\documentclass[floatfix,twocolumn,showpacs,amssymb,amsmath,prd]{revtex4}
\usepackage{graphicx}
\usepackage{dcolumn}
\usepackage{bm}

\begin{document}
\title{Symmetry and anti-symmetry of the CMB anisotropy pattern}
\author{Jaiseung Kim}
\email{jkim@nbi.dk}
\affiliation{Niels Bohr Institute \& Discovery Center, Blegdamsvej 17, DK-2100 Copenhagen, Denmark}
\author{Pavel Naselsky}
\affiliation{Niels Bohr Institute \& Discovery Center, Blegdamsvej 17, DK-2100 Copenhagen, Denmark}
\author{Martin Hansen}
\affiliation{Niels Bohr Institute \& Discovery Center, Blegdamsvej 17, DK-2100 Copenhagen, Denmark}
\date{\today}

\begin{abstract}
Given an arbitrary function, we may construct symmetric and antisymmetric functions under a certain operation. Since statistical isotropy and homogeneity of our Universe has been a fundamental assumption of modern cosmology, we do not expect any particular symmetry or antisymmetry in our Universe. Besides fundamental properties of our Universe, we may also figure our contamination and improve the quality of the CMB data products, by matching the unusual symmetries and antisymmetries of the CMB data with known contaminantions.
Noting this, we have investigated the symmetry and antisymmetry of CMB anisotropy pattern, which provides the deepest survey.
If we let the operation to be a coordinate inversion, the symmetric and antisymmetric functions have even and odd-parity respectively.
The investigation on the parity of the recent CMB data shows a large-scale odd-parity preference, which is very unlikely in the  statistical isotropic and homogeneous Universe.
We have investigated the association of the  WMAP systematics with the anomaly, but not found a definite non-cosmological cause. Additionally, we have investigated the
phase of even and odd multipole data respectively, and found the behavior distinct from each other.
Noting the odd-parity preference anomaly, we have fitted a cosmological model respectively to even and odd multipole data, and found significant parametric tension.
Besides anomalies explicitly associated with parity, there are anomalous lack of large-scale correlation in CMB data. Noting the equivalence between the power spectrum and the correlation, we have investigated the association between the lack of large-angle correlation and the odd-parity preference of the angular power spectrum. From our analysis, we find that the odd-parity preference at low multipoles is, in fact, phenomenologically identical with the lack of large-angle correlation.
\end{abstract}

\maketitle 

\section{Introduction}
In an inflationary paradigm, the CMB anisotropy pattern is expected to follow a random Gaussian distribution with the statistical isotropy and homogeneity, due to the nature of the quantum fluctuation during the cosmic inflation \citep{Cosmology_Coles,Modern_Cosmology,Foundations_Cosmology,Inflation,Cosmology}. 
Passing through the period of the inflation and cosmological evolution from very early stages till the present day, the quantum fluctuations turn into classical fluctuation, in which all information about the beginning of the inflation, the ionization history of the cosmic plasma and the formation of the large-scale structure have been well preserved. Therefore, the observation of the CMB anisotropy allows us to investigate the extreme states of matter and radiation well beyond the limit, obtainable by modern particle accelerators, and shed light on the problem of ``darkness'' of the Universe, in which the present mass density mainly consists of the cold dark matter and the dark energy.

For the past years, there have been great successes in measurement of CMB anisotropy by ground and satellite observations  \citep{WMAP7:powerspectra,WMAP7:basic_result,WMAP5:basic_result,WMAP5:powerspectra,WMAP5:parameter,ACBAR,ACBAR2008,QUaD1,QUaD2,QUaD:instrument,Planck_bluebook}. 
Since release of the data from the orbital observations \citep{Mather:CMB_T_1999,Fixen:dipole,Bennett:dipole}, the issue of statistical anisotropy and non-Gaussianity have been given very significant attention. Several hints of statistical anisotropy and non-Gaussianity have been reported \citep{WMAP7:anomaly,cold_spot1,cold_spot2,cold_spot_wmap3,Multipole_Vector1,Multipole_Vector2,Multipole_Vector3,Axis_Evil,Axis_Evil2,Axis_Evil3,lowl_WMAP13,odd_phase,PMF1_WMAP1,PMF2_WMAP1,alfven,Phase_NG,Chiang_NG,Phase_correlation,odd,odd_origin,odd_bolpol,odd_tension,odd_C,Hemispherical_asymmetry,Power_Asymmetry5}.
In particular, many of the reported anomalies are associated with low multipoles ($2\le l\le 30$) of the CMB, including the low amplitude of the quadrupole \citep{Tegmark:Alignment} and the striking alignment between the quadrupole and octupole, dubbed the `axis of evil' \citep{Tegmark:Alignment,Axis_Evil,Axis_Evil2}, and some others features of the CMB map \citep{Hemispherical_asymmetry} and the power spectrum \citep{odd,odd_origin}. These anomalies could be given two possible explanations. The first one is that statistical homogeneity and isotropy of the primordial fluctuation in general is obeyed, but we are living in a Universe, which is not typical of the ensemble Universe. The second explanation is that, at least for some range of multipoles, the properties of primordial fluctuations are in disagreement with the isotropic Gaussian Universe. 

The CMB anisotropy at low multipoles are associated with scales far beyond any existing astrophysical survey, and therefore 
CMB anomalies at low multipoles may hint new physics at unexplored large scales, including non-trivial topology of the Universe, 
broken scale invariance at large scales. On the other hand, these anomalies be simply due to non-cosmological contamination such as unaccounted astrophysical emission (e.g. the Kuiper Belt objects), unknown systematic effects and so forth. 

Recently, it was shown that some of the anomalies can be explained in terms of symmetries and antisymmetries of the CMB sky \citep{odd,odd_origin,odd_C,octupole}.
For instance, the CMB anisotropy pattern may be considered as the sum of symmetric and antisymmetric functions under the coordinate inversion. 
Equivalently, the forementioned symmetric and antisymmetric functions possess even and odd parity respectively.
Given the Gaussian Universe, we do not expect the CMB anisotropy pattern to show a particular parity preference.
However, the angular power spectrum of WMAP data shows anomalous odd-parity preference at low multipoles \citep{Universe_odd,odd,odd_origin,odd_bolpol,odd_tension}.
In this work, we are going to discuss the odd-parity preference of the WMAP data, and present our investigation on its origins.
In order to understand the nature of the odd-parity preference, we have additionally investigated the phase of even and odd multipole data respectively, and found they show features distinct from each other. The parity anomaly is explicitly associated with the angular power spectrum, which are heavily used for cosmological model fitting. 
Having noted this, we have also fitted a cosmological model respectively to even and odd multipole data set and found significant tension \citep{odd_tension}.
These parametric tensions indicate either unaccounted contamination or insufficiency of the assumed model.

One of most important element in the study of non-Gaussianity is to identify the anomalies of the common origin, whether it is cosmological or systematics. 
In particular, there have been reports on the lack of large-angle correlation, since the COBE-DMR data \citep{correlation_COBE,WMAP1:Cosmology,correlation_Copi1,correlation_Copi2,lowl_anomalies,lowl_bias}.
Recently, we have shown that the lack of large-angle correlation is phenomenologically identical with the odd-parity anomaly of the CMB power spectrum. 
Besides them, we may understand the low quadrupole power as a part of odd-parity preference at low multipoles \citep{Tegmark:Alignment,odd,odd_origin}.
Even though it still leaves the fundamental question on its origin unanswered, the association between seemingly distinct anomalies will help the investigation on the underlying origin.

The outline of the paper is the following. 
In Section \ref{asymmetry}, we discuss the anomalous odd-parity preference of the WMAP data.
In Section \ref{phase}, we investigate the phase of the even and odd multipole data respectively, and discuss its result.
In Section \ref{S3}, we investigate the octupole component of CMB anisotropy and discuss some anomalous feature.
In Section \ref{cosmomc}, we show there is a significant parametric tension between the cosmological models, when fitted to the even or odd low multipole data respectively.
In Section \ref{correlation} and \ref{association}, we discuss the lack of correlation of WMAP data at large and small angles, and show the odd-parity preference at low multipoles is phenomenologically identical with the lack of the large-angle correlation.
Finally, in Section \ref{discussion}, we discuss the findings and draw our conclusions.

\section{Parity asymmetry of the WMAP data}
\label{asymmetry}
The CMB temperature anisotropy over a whole-sky is conveniently decomposed in terms of spherical harmonics $Y_{lm}(\theta,\phi)$ as follows:
\begin{eqnarray}
T(\hat{\mathbf n})=\sum_{lm} a_{lm}\,Y_{lm}(\hat{\mathbf n}), \label{T_expansion}
\end{eqnarray}
where $a_{lm}$ is a decomposition coefficient, and $\hat{\mathbf n}$ is a sky direction.
Decomposition coefficients are related to primordial perturbation as follows:
\begin{eqnarray}
a_{lm}&=&4\pi (-\imath)^l \int \frac{d^3\mathbf k}{(2\pi)^3} \Phi(\mathbf k)\,g_{l}(k)\,Y^*_{lm}(\hat {\mathbf k}),\label{alm}
\end{eqnarray} 
where $\Phi(\mathbf k)$ is primordial perturbation in Fourier space, and $g_{l}(k)$ is a radiation transfer function.
For a Gaussian model for primordial perturbation, decomposition coefficients satisfy the following statistical properties:
\begin{eqnarray} 
\langle a_{lm} \rangle &=& 0, \\
\langle a^*_{lm} a_{l'm'} \rangle &=& C_l\,\delta_{ll'}\delta_{mm'} \label{alm_cor},
\end{eqnarray}
where $\langle\ldots\rangle$ denotes the average over the ensemble of universes.
Given a standard cosmological model, Sach-Wolf plateau is expected at low multipoles  \citep{Modern_Cosmology}:
$l(l+1) C_l\sim \mathrm{const}$.

\begin{figure}[htb!]
\centering\includegraphics[scale=.5]{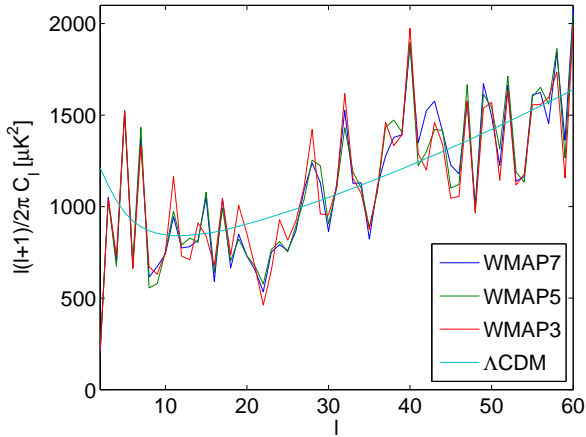}
\caption{CMB power spectrum: WMAP 7 year data (blue), WMAP 5 year data (green) and WMAP 3 year data (red), $\Lambda$CDM model (cyan)}
\label{Cl_cut}
\end{figure}

From the CMB anisotropy, we may construct a symmetric and antisymmetric function under the coordinate inversion $\mathbf n\rightarrow -\mathbf n$:
\begin{eqnarray} 
T^+(\hat{\mathbf n})&=&\frac{T(\hat{\mathbf n})+T(-\hat{\mathbf n})}{2},\\
T^-(\hat{\mathbf n})&=&\frac{T(\hat{\mathbf n})-T(-\hat{\mathbf n})}{2}.
\end{eqnarray}
In other words, $T^+(\hat{\mathbf n})$) and $T^-(\hat{\mathbf n})$ have even and odd-parity.
Taking into account the parity property of spherical harmonics $Y_{lm}(\hat{\mathbf n})=(-1)^l\,Y_{lm}(-\hat{\mathbf n})$ \citep{Arfken},
we may easily show
\begin{eqnarray} 
T^+(\hat{\mathbf n})&=&\sum_{lm} a_{lm}\,Y_{lm}(\hat{\mathbf n})\,\cos^2\left(\frac{l\pi}{2}\right), \label{T_even}\\
T^-(\hat{\mathbf n})&=&\sum_{lm} a_{lm}\,Y_{lm}(\hat{\mathbf n})\,\sin^2\left(\frac{l\pi}{2}\right), \label{T_odd}
\end{eqnarray}
where $n$ is an integer.
Therefore, significant power asymmetry between even and odd multipoles may be interpreted as a preference for a particular parity of the anisotropy pattern. 
Hereafter, we will denote a preference for particular parity by `parity asymmetry'. 
In Fig. \ref{Cl_cut}, we show the WMAP 7 year, 5 year, 3 year data and the WMAP concordance model  \citep{WMAP7:powerspectra,WMAP5:powerspectra,WMAP3:temperature,WMAP5:Cosmology,WMAP7:Cosmology}. 
From Fig. \ref{Cl_cut}, we may see that the power spectrum of WMAP data at even multipoles tend to be lower than those at neighboring odd multipoles. 
\begin{figure}[htb]
\centering\includegraphics[scale=.5]{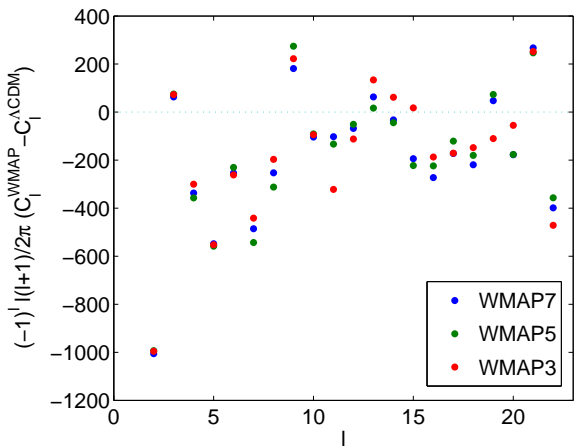}
\caption{$(-1)^l\times$ difference between WMAP power spectrum data and $\Lambda$CDM model}
\label{delta}
\end{figure}
In Fig. \ref{delta}, we show $(-1)^l l(l+1)/2\pi\:(C^{\mathrm{WMAP}}_l-C^{\Lambda\mathrm{CDM}}_l)$ for low multipoles.
Since we expect random scattering of data points around a theoretical model, we expect the distribution of dots in Fig. \ref{delta} to be scattered around the both side of zero. However, there are only 5 points of positive values among 22 points in the case of WMAP7 or WMAP5 data.
Therefore, we may see that there is the tendency of power deficit (excess) at even (odd) multipoles, compared with the $\Lambda$CDM model.  
Taking into account $l(l+1) C_l\sim \mathrm{const}$, we may consider the following quantities:
\begin{eqnarray} 
P^{+} &=& \sum^{l_{\mathrm{max}}}_{l=2} \,\cos^2\left(\frac{l\pi}{2}\right)\, \frac{l(l+1)}{2\pi} \: C_l,\\
P^{-} &=& \sum^{l_{\mathrm{max}}}_{l=2} \,\sin^2\left(\frac{l\pi}{2}\right)\, \frac{l(l+1)}{2\pi} \: C_l,
\end{eqnarray}
where $P^{+}$ and $P^{-}$ are the sum of $l(l+1)/2\pi \: C_l$ for even and odd multipoles respectively. 
Therefore, the ratio $P^{+}/P^{-}$ is associated with the degree of the parity asymmetry, where the lower value of $P^+/P^-$ indicates odd-parity preference, and vice versa. 

\begin{figure}[htb]
\centering\includegraphics[scale=.5]{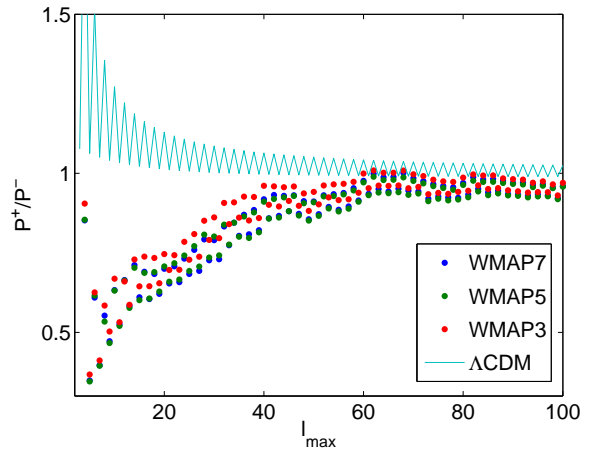}
\caption{$P^+/P^-$ of WMAP data and $\Lambda$CDM}
\label{P_ratio}
\end{figure}
In Fig. \ref{P_ratio}, we show the $P^+/P^-$ of WMAP data, and a $\Lambda$CDM model for various $l_{\mathrm{max}}$.
As shown in Fig. \ref{P_ratio}, $P^+/P^-$ of WMAP data are far below theoretical values. Though the discrepancy is largest at lowest $l_{\mathrm{max}}$, its statistical significance is not necessarily high for low $l$, due to associated statistical fluctuation.
In order to make a rigorous assessment on its statistical significance at low $l$, we compared $P^+/P^-$ of WMAP data with that of simulation.
We have produced $10^4$ simulated CMB maps of HEALPix Nside=8 and Nside=512 respectively, via map synthesis with $a_{lm}$ randomly drawn from Gaussian $\Lambda$CDM model.
We have degraded the WMAP processing mask (Nside=16) to Nside=8, and set pixels to zero, if any of their daughter pixels is zero. 
After applying the mask, we have estimated power spectrum $2\le l\le 23$ from simulated cut-sky maps (Nside=8) by a pixel-based maximum likelihood method \citep{WMAP7:powerspectra,Bond:likelihood,hybrid_estimation}. 
At the same time, we have applied the WMAP team's KQ85 mask to the simulated maps (Nside=512), and estimated power spectrum $2\le l\le 1024$  by pseudo $C_l$ method \citep{pseudo_Cl,MASTER}. In the simulation, we have neglected instrument noise, since the signal-to-noise ratio of the WMAP data is quite high at multipoles of interest (i.e. $l\le 100$) \citep{WMAP7:powerspectra,WMAP7:basic_result}. Using the low $l$ estimation by pixel-maximum likelihood method and high $l$ estimation by pseudo $C_l$ method, we have computed $P^+/P^-$ respectively for various multipole ranges $2\le l \le l_{\mathrm{max}}$, and compared $P^+/P^-$ of the WMAP data with simulation.
\begin{figure}[htb]
\centering\includegraphics[scale=.5]{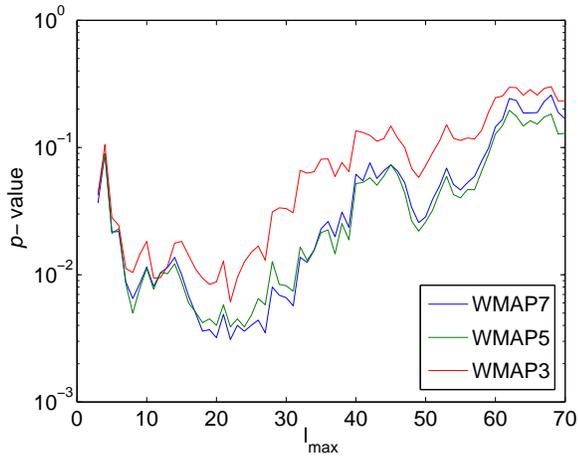}
\caption{Probability of getting $P^+/P^-$ as low as WMAP data for multipole range $2\le l\le{l_\mathrm{max}}$.}
\label{P}
\end{figure}
In Fig. \ref{P}, we show $p$-value of WMAP7, WMAP5 and WMAP3 respectively for various $l_{\mathrm{max}}$, where $p$-value denotes fractions of simulations as low as $P^+/P^-$ of the WMAP data. As shown in Fig. \ref{P}, the parity asymmetry of WMAP7 data at multipoles ($2\le l\le 22$) is most anomalous, where $p$-value is $0.0031$.
As shown in Fig. \ref{P}, the statistical significance of the parity asymmetry (i.e. low $p$-value) is getting higher, when we increase $l_{\mathrm{max}}$ up to 22.
Therefore, we may not attribute the odd parity preference simply to the low quadrupole power, and find it rather likely that the low quadrupole power is not an isolated anomaly, but shares an origin with the odd parity preference. 

\begin{figure}[htb]
\centering\includegraphics[scale=.5]{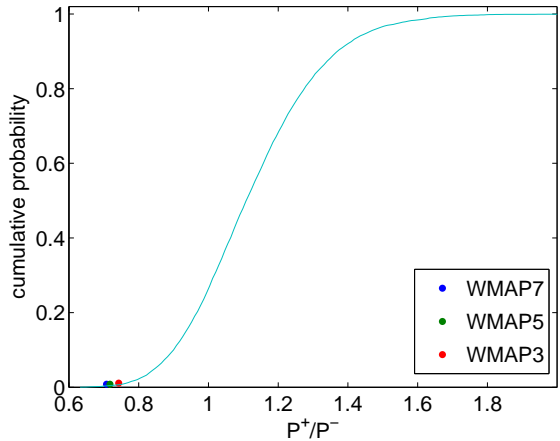}
\caption{Parity asymmetry at multipoles ($2\le l\le 22$): cumulative distribution of $P^+/P^-$ for $10^4$ simulated maps (cyan), $P^+/P^-$ of WMAP7 (blue), WMAP5 (green) and WMAP3 (red)}
\label{hist}
\end{figure}

\begin{table}[htb]
\centering
\caption{the parity asymmetry of WMAP data ($2\le l\le 22$)}
\begin{tabular}{ccc}
\hline\hline 
data &  $P^+/P^-$  &  $p$-value\\
\hline
WMAP7  &  0.7076  & 0.0031\\
WMAP5  &  0.7174 &  0.0039\\
WMAP3  &  0.7426 & 0.0061\\
\hline
\end{tabular}
\label{pvalue}
\end{table}
In Table \ref{pvalue}, we summarize $P^+/P^-$ and $p$-values of WMAP7, WMAP5 and WMAP3 for $l_{\mathrm{max}}=22$. As shown in Fig. \ref{P} and Table \ref{pvalue}, the odd-parity preference of WMAP7 is most anomalous, while WMAP7 data are believed to have more accurate calibration and less foreground contamination than earlier releases.  \citep{WMAP7:powerspectra,WMAP7:basic_result,WMAP5:basic_result,WMAP5:powerspectra,WMAP5:beam}. 
In Fig. \ref{hist}, we show cumulative distribution of $P^+/P^-$ for $10^4$ simulated maps. The values corresponding to $P^+/P^-$ of WMAP data are marked as dots.

In the absence of strong theoretical grounds for the parity asymmetry ($2\le l\le 22$),
we have to take into account our posteriori choice on $l_{\mathrm{max}}$, which might have enhanced the statistical significance.
However, as shown in Fig. \ref{P}, the odd-parity preference exists for various values of $l_\mathrm{max}$.
Therefore, the statistical enhancement by our posterior choice on  $l_{\mathrm{max}}$ is not significant.

\subsection{cosmological or non-cosmological?}
\begin{figure}[htb]
\centering
\includegraphics[scale=.5]{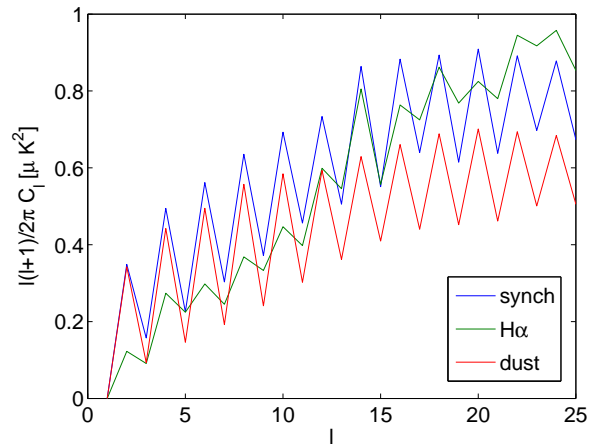}
\caption{the power spectra of the templates (synchrotron, H$\alpha$, dust): plotted with arbitrary normalization.}
\label{template}
\end{figure}

In the WMAP data, there are non-cosmological contamination such as asymmetric beams, instrument noise, foreground and cut-sky effect, which might be responsible for the discussed anomaly.
First of all, there are contamination from galactic and extragalactic foregrounds.
In order to reduce foreground contamination, the WMAP team have subtracted diffuse foregrounds by template-fitting, and masked the regions that cannot be cleaned reliably.
For foreground templates (dust, free-free emission and synchrotron), the WMAP team used dust emission ``Model 8", H$\alpha$ map, and the difference between K and Ka band maps  \citep{WMAP5:basic_result,WMAP7:fg,Dust_Extrapolation,Finkbeiner_H_alpha,WMAP1:fg}. 
In Fig. \ref{template}, we show the power spectrum of templates. 
As shown in Fig. \ref{template}, templates show strong even parity preference, which is opposite to that of the WMAP power spectrum data.
Therefore, one might wrongly attribute the odd-parity preference of WMAP data to over-subtraction by templates.
Consider spherical harmonic coefficients of a foreground-reduced map:
\begin{eqnarray} 
a^{\mathrm{obs}}_{lm}=a^{\mathrm{cmb}}_{lm}+a^{\mathrm{fg}}_{lm}-b\,a^{\mathrm{tpl}}_{lm},
\end{eqnarray}
where $a^{\mathrm{obs}}_{lm}$, $a^{\mathrm{fg}}_{lm}$ and $b\,a^{\mathrm{tpl}}_{lm}$ correspond to a foreground-cleaned map, a foreground and
a template with a fitting coefficient $b$.
For simplicity, we consider only a single foreground component, but the conclusion is equally valid for multi-component foregrounds.
Since there is no correlation between foregrounds and CMB, the observed power spectrum is given by:
\begin{eqnarray} 
C^{\mathrm{obs}}_l\approx C^{\mathrm{cmb}}_l+ \langle \left|a^{\mathrm{fg}}_{lm}  -b\, a^{\mathrm{tpl}}_{lm} \right|^2\rangle. \label{Cl_obs}
\end{eqnarray}
As shown Eq. \ref{Cl_obs},  the parity preference should follow that of templates (i.e. even parity preference), because of the second term, provided templates are good tracers of foregrounds (i.e. $a^{\mathrm{fg}}_{lm}/a^{\mathrm{tpl}}_{lm}\approx \mathrm {const}$). Nevertheless, Eq. \ref{Cl_obs} may make a bad approximation for lowest multipoles, because the cross term $\sum_m \mathrm{Re}[a^{\mathrm{cmb}}_{lm} (a^{\mathrm{fg}}_{lm}  -b\, a^{\mathrm{tpl}}_{lm})^*]$ may not be negligible. 
Besides that, our argument and the template-fitting method itself fail, if templates are not good tracers of foregrounds. 
In order to investigate these issues, we have resorted to simulation in combination with WMAP data.
Noting the WMAP power spectrum is estimated from foreground-reduced V and W band maps, we have produced simulated maps as follows:
\begin{eqnarray} 
T(\hat{\mathbf n})=T_{\mathrm{cmb}}(\hat{\mathbf n})+(V(\hat{\mathbf n})-W(\hat{\mathbf n}))/2,  \label{residual_fg2}
\end{eqnarray}
where $V(\hat{\mathbf n})$ and $W(\hat{\mathbf n})$ are foreground-reduced V and W band maps of WMAP data. Note that the second term on the right hand side mainly contains residual foregrounds.
Just as cut-sky simulation described in Sec. \ref{asymmetry}, we have applied a foreground mask to the simulated maps, and estimated the power spectrum from cut-sky by a pixel-based maximum likelihood method.
\begin{figure}[htb]
\centering
\includegraphics[scale=.5]{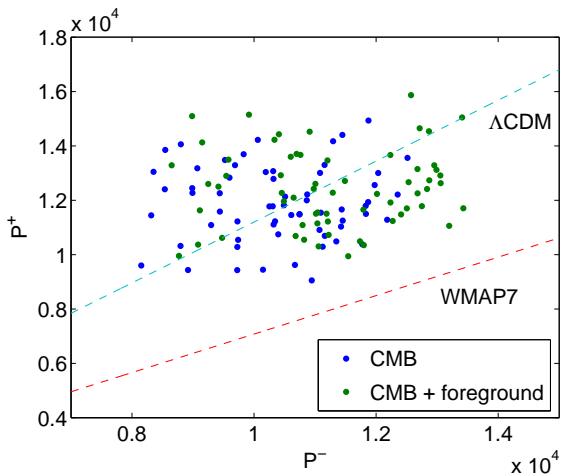}
\caption{the parity asymmetry in the presence of residual foregrounds (V$-$W): Dashed lines are plotted with slopes corresponding to $P^+/P^-$ of $P^+/P^-$ of $\Lambda$CDM (cyan), WMAP7 data (red).}
\label{residual2}
\end{figure} 
In Fig. \ref{residual2}, we show $P^+$ and $P^-$ values estimated from simulations. 
For comparison, we have included simulations without residual foregrounds, and dashed lines of a slope corresponding to $P^+/P^-$ of $\Lambda$CDM model and WMAP7 data. As shown in Fig. \ref{residual2}, the $P^+/P^-$ of simulations in the presence of residual foregrounds do not show anomalous odd-parity preference of WMAP data.
Considering Eq. \ref{Cl_obs} and simulations, we find it difficult to attribute the odd-parity preference to residual foreground.
There also exist contamination from unresolved extragalactic point sources \citep{WMAP3:temperature}. 
However, point sources follow Poisson distribution with little departure \citep{Tegmark:Foreground}, and therefore are unlikely to possess odd-parity preference.
Besides that, point sources at WMAP frequencies are subdominant on large angular scales (low $l$) \citep{WMAP3:temperature,WMAP7:fg,Tegmark:Foreground,WMAP5:fg}.
Though we have not found association of foregrounds with the anomaly, we do not completely rule out residual foreground, due to our limited knowledge on residual foregrounds.

The WMAP team have masked the region that cannot be reliably cleaned by template fitting, and estimated CMB power spectrum from sky data outside the mask  \citep{WMAP7:powerspectra,WMAP5:powerspectra,WMAP3:temperature,WMAP7:fg}. 
Even though we have properly taken into account the cut-sky effect in the $p$-value estimation, we have investigated the WMAP team's Internal Linear Combination (ILC) map, which is expected to provide a reliable estimate of CMB signal over whole-sky on angular scales larger than $10^\circ$  \citep{WMAP3:temperature,WMAP7:fg,WMAP5:fg}.
\begin{figure}[htb]
\centering
\includegraphics[scale=.5]{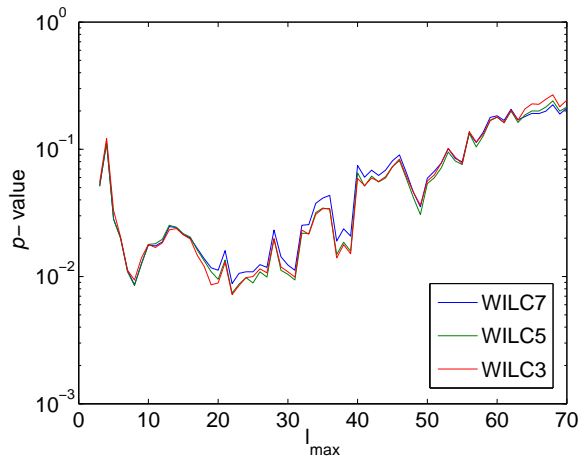}
\caption{Probability of getting $P^+/P^-$ as low as the ILC 7 year, 5 year, and 3 year map at multipole range $2\le l\le{l_\mathrm{max}}$}
\label{P_wilc}
\end{figure}
We have compared $P^+/P^-$ of the ILC maps with whole-sky simulations.
In Fig. \ref{P_wilc}, we show $p$-values of the ILC maps respectively for various $l_{\mathrm{max}}$.
As shown in Fig. \ref{P_wilc}, the odd-parity preference of ILC maps is most anomalous for $l_{\mathrm{max}}=22$ as well.
In Table \ref{pvalue_wilc}, we summarize $P^+/P^-$ and $p$-values for $l_{\mathrm{max}}=22$. 
\begin{table}[htb]
\centering
\caption{the parity asymmetry of WMAP ILC maps ($2\le l\le 22$)}
\begin{tabular}{ccc}
\hline\hline 
data &  $P^+/P^-$  &  $p$-value\\
\hline
ILC7  &  0.7726  & 0.0088\\
ILC5  &  0.7673 &  0.0074\\
ILC3  &  0.7662 & 0.0072\\
\hline
\end{tabular}
\label{pvalue_wilc}
\end{table}
As shown in Fig. \ref{P_wilc} and Table \ref{pvalue_wilc}, we find anomalous odd-parity preference exits in whole-sky CMB maps as well.
Therefore, we find it difficult to attribute the anomaly to cut-sky effect.

\begin{figure}[htb]
\centering
\includegraphics[scale=.5]{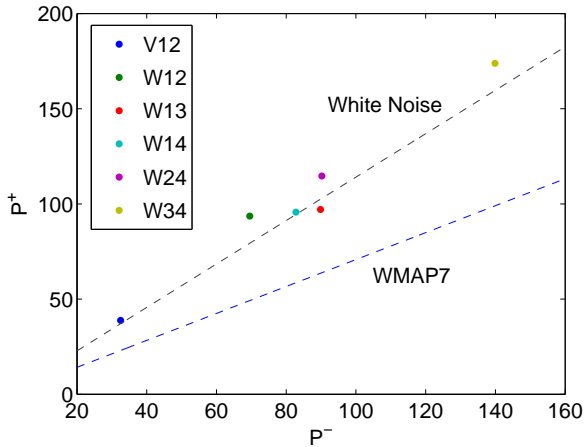}
\caption{the parity asymmetry of the WMAP noise: 
The dots denotes ($P^+$,$P^-$) of noise maps, and alphanumeric values in the legend denote the frequency band and the pair of D/A channels used. Two dashed lines are plotted with the slope corresponding to $P^+/P^-$ of white noise and WMAP7 data respectively.}
\label{noise}
\end{figure}

There are instrument noise in the WMAP data.
Especially, 1/f noise, when coupled with WMAP scanning pattern, may result in less accurate measurement at certain low multipoles  \citep{WMAP3:temperature,Detection_Light,WMAP1:processing}. 
In order to investigate the association of noise with the anomaly, we have produced noise maps of WMAP7 data by subtracting one Differencing Assembly (D/A) map from another D/A data of the same frequency channel. 
In Fig. \ref{noise}, we show $P^+$ and $P^-$ values of the noise maps.
As shown in Fig. \ref{noise}, the noise maps do not show odd-parity preference, but their $P^+/P^-$ ratios are consistent with that of white noise (i.e. $C_l=\mathrm{const.}$).
Besides that, the signal-to-noise ratio of WMAP temperature data is quite high at low multipoles (e.g S/N$\sim$ 100 for $l=30$)  \citep{WMAP3:temperature,WMAP5:beam,WMAP1:processing}. 
Therefore, we find that instrument noise, including 1/f noise, is unlikely to be the cause of the odd-parity preference. 

\begin{figure}[htb]
\centering
\includegraphics[scale=.25]{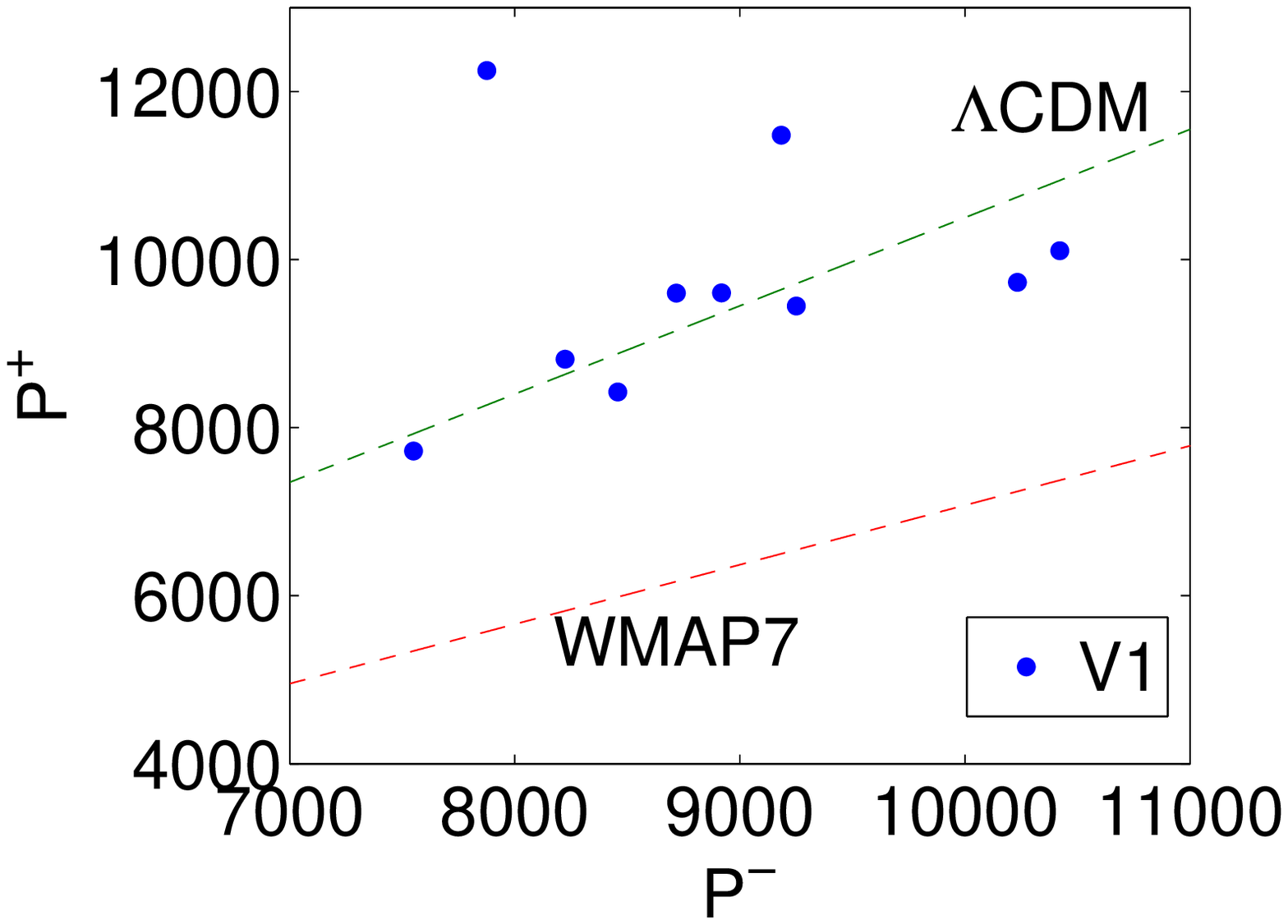}
\includegraphics[scale=.25]{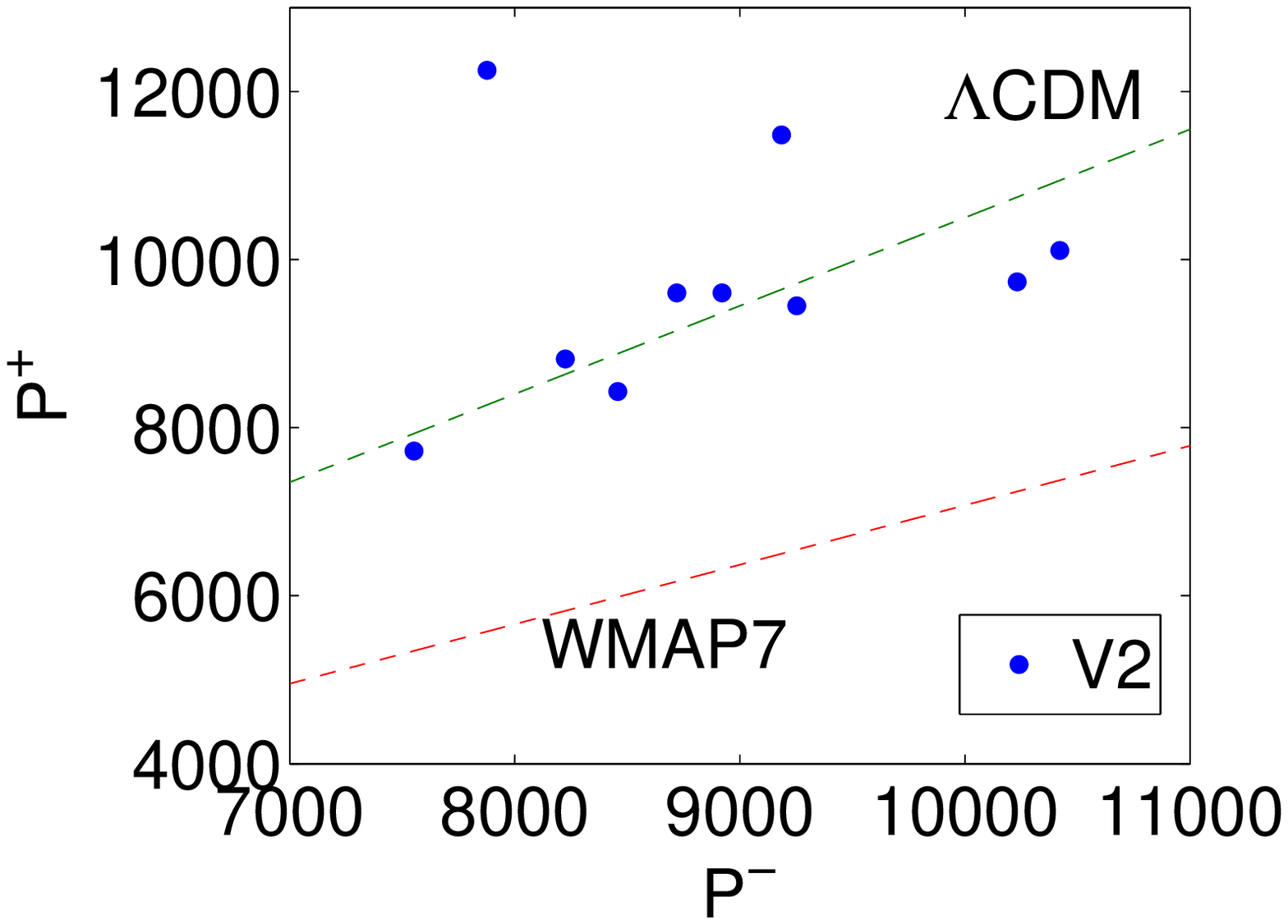}
\includegraphics[scale=.25]{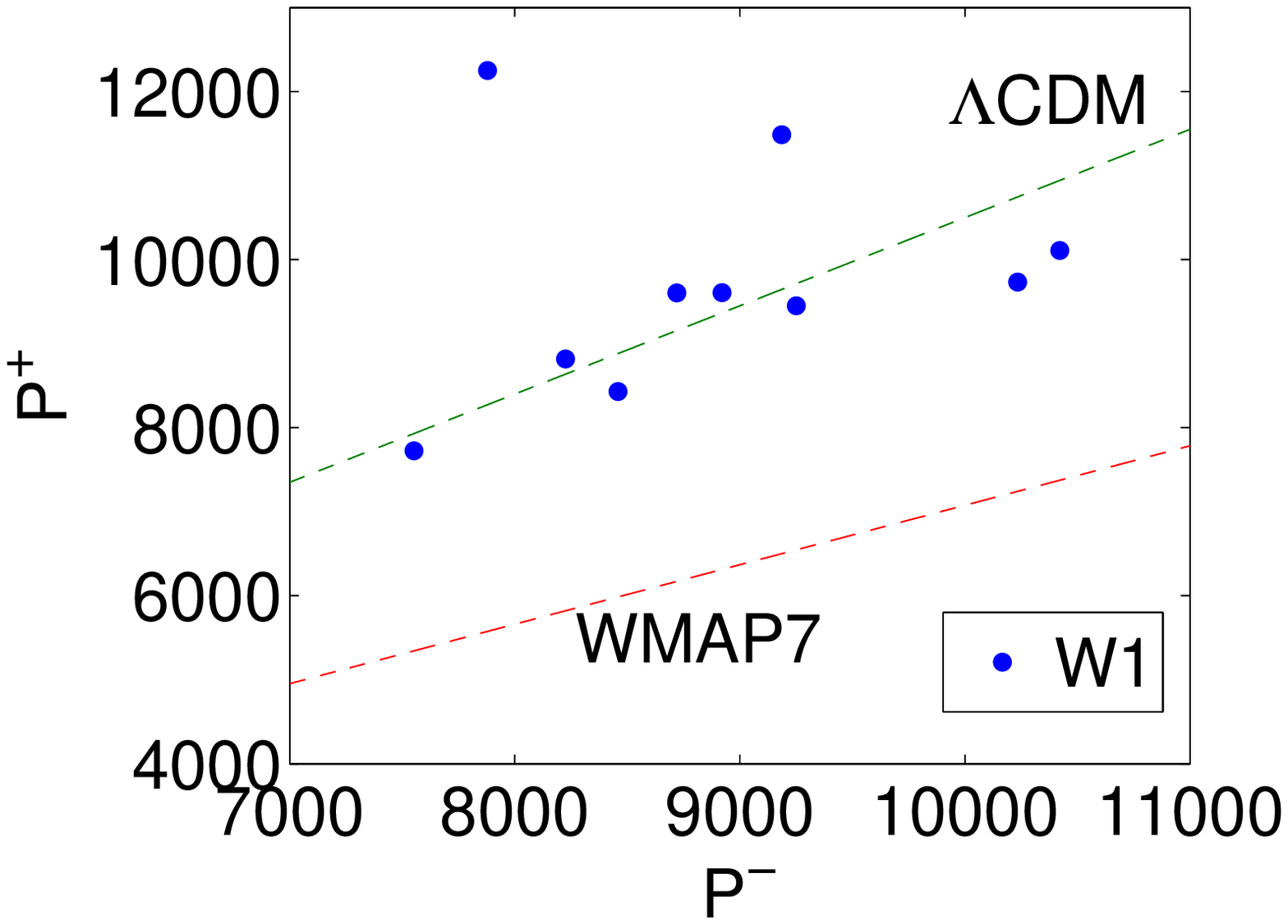}
\includegraphics[scale=.25]{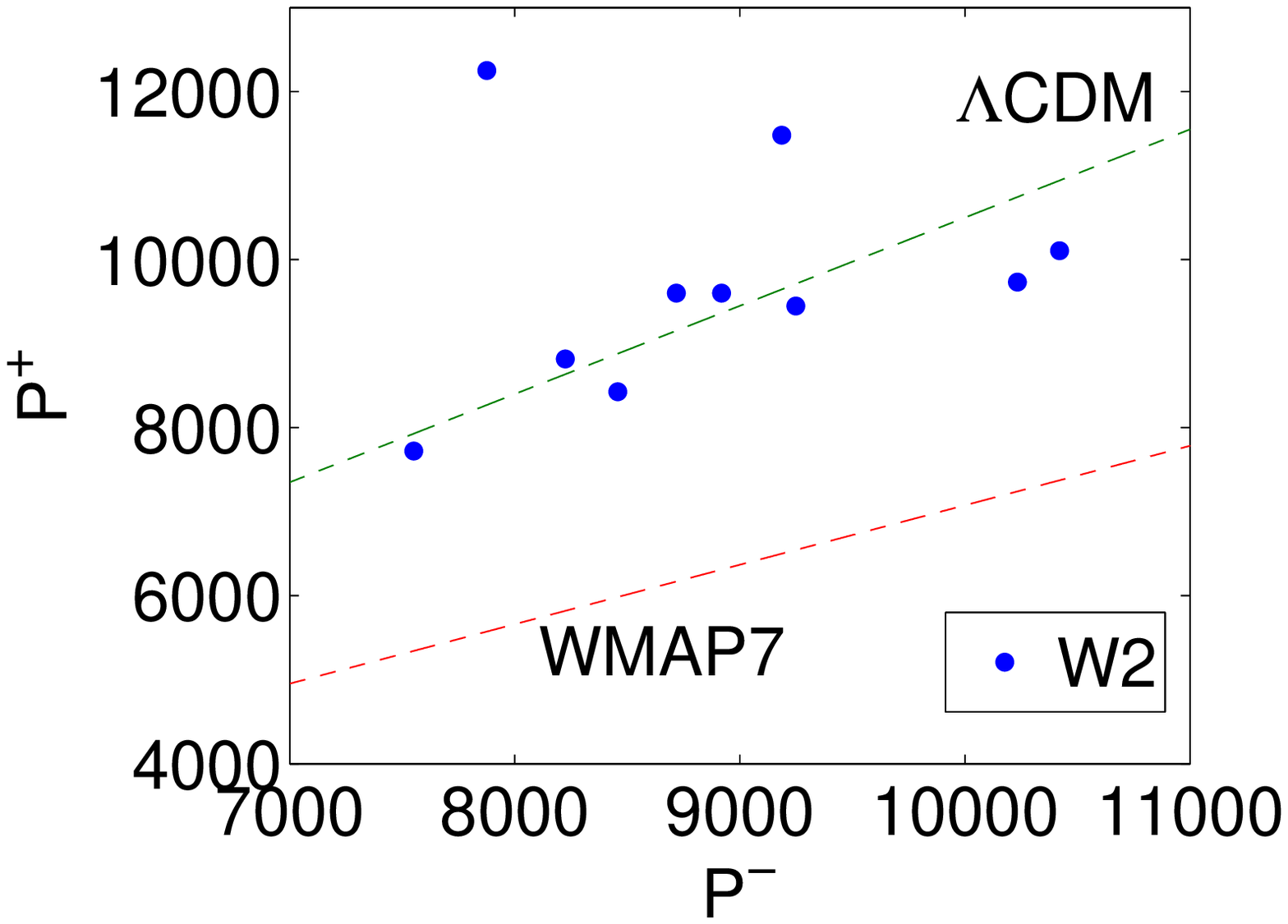}
\includegraphics[scale=.25]{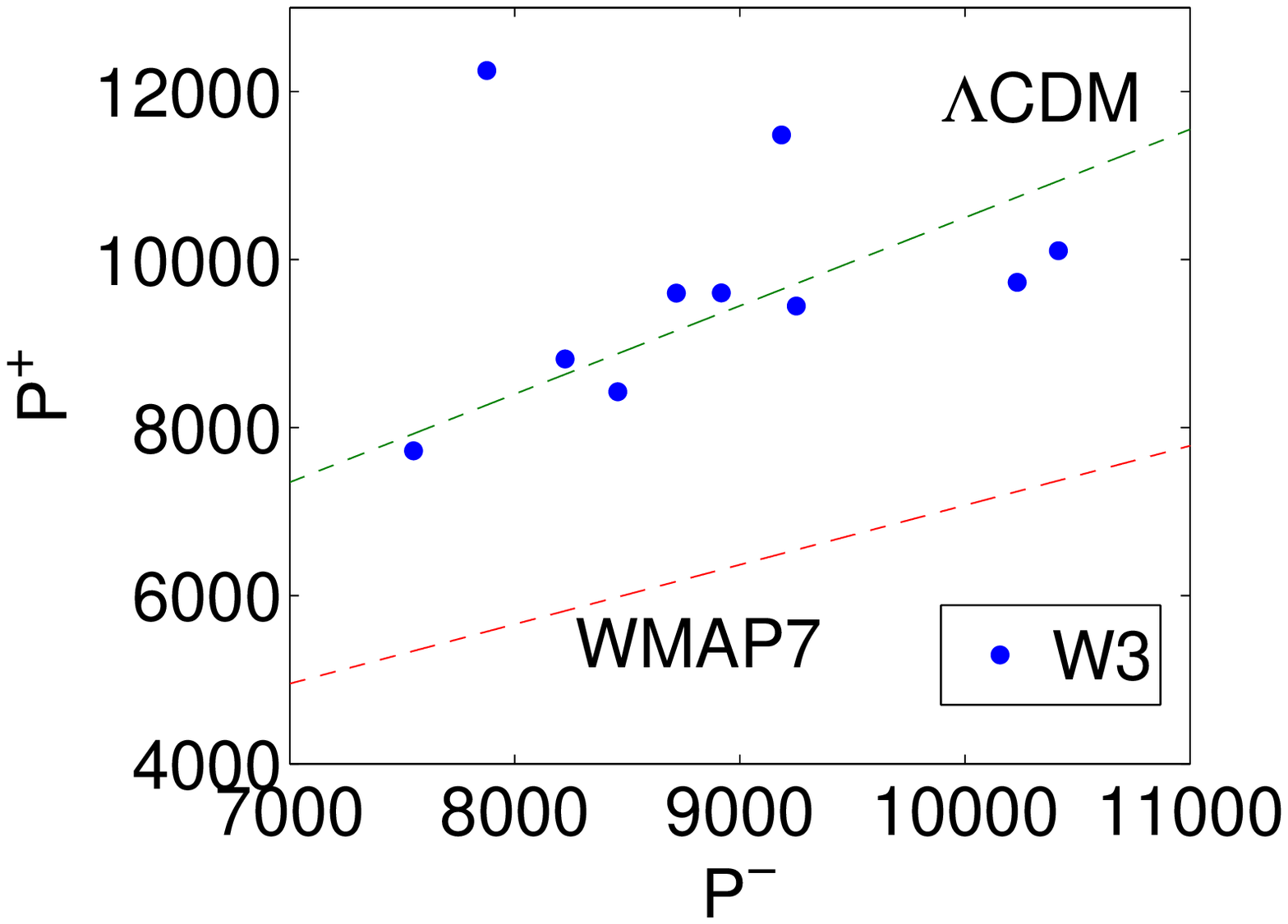}
\includegraphics[scale=.25]{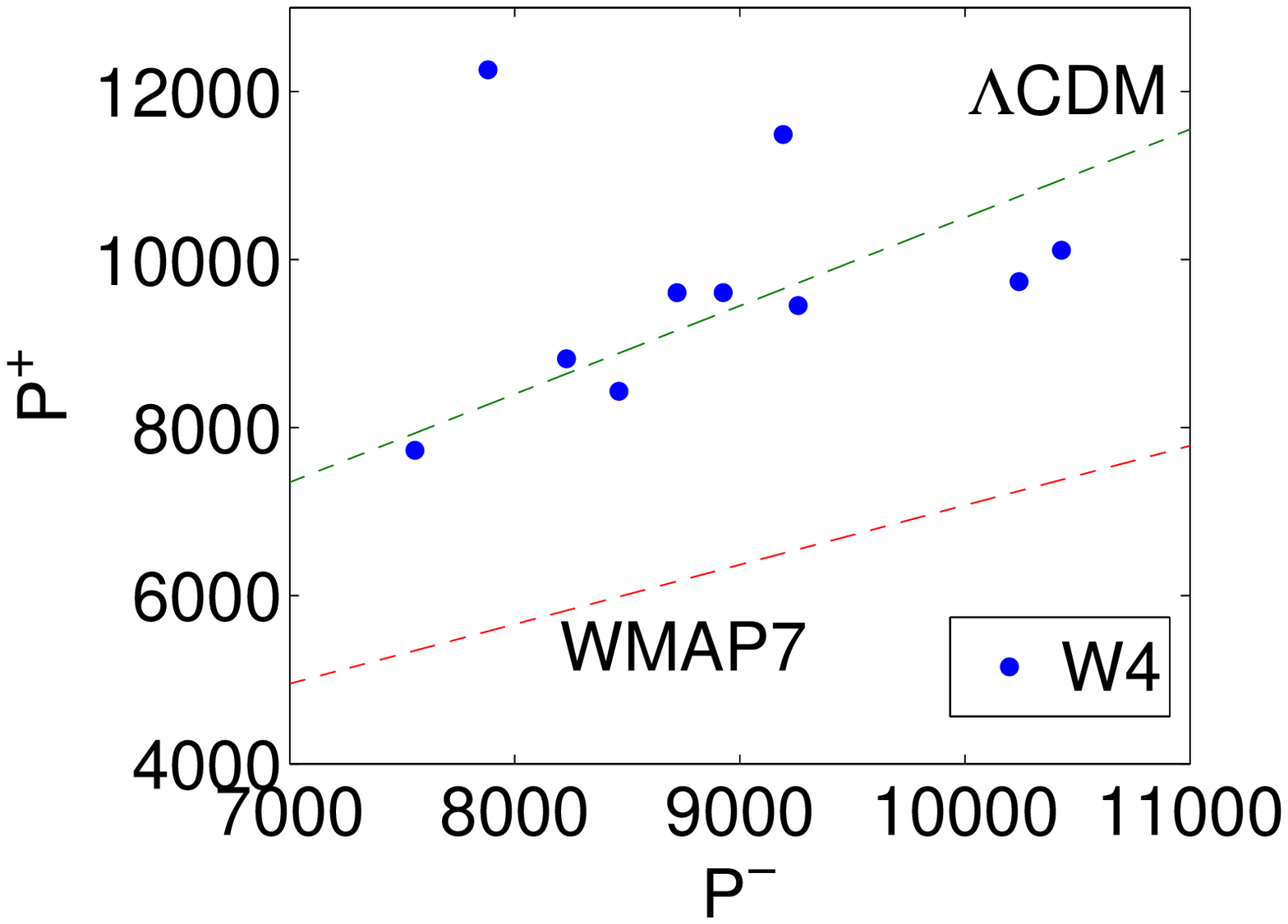}
\caption{the parity asymmetry in the presence of beam asymmetry: 
The dots denotes ($P^+$,$P^-$) of CMB maps simulated with asymmetric beams. 
The dashed lines are plotted with slopes corresponding to the $P^+/P^-$ of $\Lambda$CDM model (red) and WMAP7 data (green) respectively.
The alphanumeric values at the lower right corner denote the frequency band and D/A channel.}
\label{beam}
\end{figure}

The shape of the WMAP beams are slightly  asymmetric \citep{WMAP5:beam,WMAP3:beam,asymmetric_beam}, while
the WMAP team have assumed symmetric beams in the power spectrum estimation \citep{WMAP7:powerspectra,WMAP5:powerspectra,WMAP5:beam,WMAP3:beam}.
We have investigated the association of beam asymmetry with the anomaly, by using simulated maps provided by \citep{asymmetric_beam}.
The authors have produced 10 simulated maps for each frequency and Differencing Assembly (D/A) channels, where the detailed shape of the WMAP beams and the WMAP scanning strategy are taken into account \citep{asymmetric_beam}.
From simulated maps, we have estimated $P^+$ and $P^-$, where we have compensated for beam smoothing purposely by the WMAP team's beam transfer function (i.e. symmetric beams).
In Fig. \ref{beam}, we show $P^+$ and $P^-$ values of the simulated maps, and the dashed lines of a slope corresponding to $P^+/P^-$ of $\Lambda$CDM and WMAP7 data respectively.
As shown in Fig. \ref{beam}, we do not observe the odd-parity preference of WMAP data in simulated maps.
Therefore, we find it hard to attribute the odd-parity preference to asymmetric beams.

Besides contamination discussed so far, there are other sources of contamination such as far sidelobe pickup, and  so on.
In order to investigate these effects, we have resorted to simulation produced by the WMAP team.
According to the WMAP team, time-ordered data (TOD) have been simulated with realistic noise, thermal drifts in instrument gains and baselines,
smearing of the sky signal due to finite integration time, transmission imbalance, and far-sidelobe beam pickup.
Using the same data pipeline used for real data, the WMAP team have processed simulated TOD, and produced maps for each differencing assembly and each single year observation year.
\begin{figure}[htb]
\centering
\includegraphics[scale=.5]{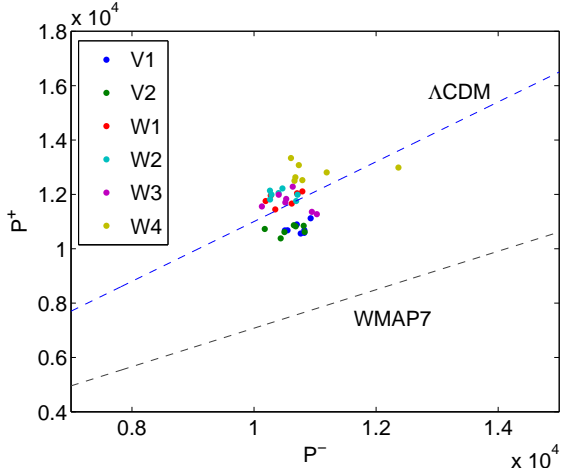}
\caption{ $P^+$ and $P^-$ of the WMAP team's simulation for V and W band data}
\label{sim}
\end{figure}
In Fig. \ref{sim}, we show the $P^+$ and $P^-$ of the simulated maps, where the power spectrum estimation is made from cut-sky by a pixel-based likelihood method. 
As shown in Fig. \ref{sim}, all points are far above $P^+/P^-$ of WMAP7, and agree with $\Lambda$CDM model.
Therefore, we do not find definite association of the parity asymmetry with known systematics effects. 

As discussed, we are unable to find a definite non-cosmological cause of the anomaly.
Therefore, we are going to take the WMAP power spectrum at face values, and consider a possible cosmological origin.
Topological models including multi-connected Universe and Bianchi $VII$ model have been proposed to explain the cold spot or low quadrupole power  \citep{low_quadrupole,spherical_tessellation,Land_Bianchi}.
However, the topological models do not produce the parity asymmetry, though some of them, indeed, predict low quadrupole power.
Trans-Planckian effects and some inflation models predict oscillatory features in primordial power spectrum  \citep{Inflation,Inflation_Planckian_problem,Inflation_Planckian_spectra,Inflation_Planckian_note,Inflation_Planckian_estimate,CMB_Planckian_signature,WMAP_oscillation,Inflation_Planckian,Inflation_initial,Planckian_Astrophysics,CMB_Planckian_observation,WMAP3:parameter}. 
However, oscillatory or sharp features in primordial power spectrum are smeared out in translation to the CMB power spectrum \citep{WMAP7:anomaly}.
Besides, reconstruction of primordial power spectrum and investigation on features show that primordial power spectrum is close to a featureless power-law spectrum \citep{WMAP7:powerspectra,WMAP5:Cosmology,WMAP7:Cosmology,WMAP3:parameter,power_recon,power_svd,power_features}.
Therefore, we find it difficult to attribute the anomaly to trans-Planckian effect or extended inflation models.
We will consider what the odd-parity preference imply on primordial perturbation $\Phi(\mathbf k)$. Using Eq. \ref{alm}, we may show the decomposition coefficients of CMB anisotropy are given by: 
\begin{eqnarray*}
a_{lm}&=&\frac{(-\imath)^l}{2\pi^2} \int\limits^{\infty}_0 dk  \int\limits^{\pi}_0 d\theta_{\mathbf k} \sin\theta_{\mathbf k}  \int\limits^{2\pi}_0 d\phi_{\mathbf k}\,\Phi(\mathbf k)\,g_{l}(k)\,Y^*_{lm}(\hat{\mathbf k}),\nonumber\\
&=&\frac{(-\imath)^l}{2\pi^2} \int\limits^{\infty}_0 dk  \int\limits^{\pi}_0 d\theta_{\mathbf k} \sin\theta_{\mathbf k}\int\limits^{\pi}_0 d\phi_{\mathbf k}\,g_{l}(k)\times\\
&&\left(\Phi(\mathbf k)\,Y^*_{lm}(\hat {\mathbf k}) + \Phi(-\mathbf k)\,Y^*_{lm}(-\hat {\mathbf k}) \right),\nonumber\\
&=&\frac{(-\imath)^l}{2\pi^2} \int\limits^{\infty}_0 dk  \int\limits^{\pi}_0 d\theta_{\mathbf k} \sin\theta_{\mathbf k}\int\limits^{\pi}_0 d\phi_{\mathbf k}\,g_{l}(k) Y^*_{lm}(\hat {\mathbf k})\times\\
&&\left(\Phi(\mathbf k)+(-1)^l \Phi^*(\mathbf k)\right),
\end{eqnarray*}
where we used the reality condition $\Phi(-\mathbf k)= \Phi^*(\mathbf k)$ and $Y_{lm}(\hat{-\mathbf n})=(-1)^l\,Y_{lm}(\hat{\mathbf n})$.
Using Eq. \ref{alm2},  it is trivial to show, for the odd number multipoles $l=2n-1$,
\begin{eqnarray}
\lefteqn{a_{lm}=}\label{alm2}\\
&&-\frac{(-\imath)^{l-1}}{\pi^2} \int\limits^{\infty}_0 dk  \int\limits^{\pi}_0 d\theta_{\mathbf k} \sin\theta_{\mathbf k}\int\limits^{\pi}_0 d\phi_{\mathbf k}\,g_{l}(k) Y^*_{lm}(\hat {\mathbf k})\,\mathrm{Im}[\Phi(\mathbf k)],\nonumber
\end{eqnarray}
and, for even number multipoles $l=2n$,
\begin{eqnarray}
\lefteqn{a_{lm}=}\label{alm3}\\
&&\frac{(-\imath)^l}{\pi^2} \int\limits^{\infty}_0 dk  \int\limits^{\pi}_0 d\theta_{\mathbf k} \sin\theta_{\mathbf k}\int\limits^{\pi}_0 d\phi_{\mathbf k}\,g_{l}(k) Y^*_{lm}(\hat {\mathbf k})\,\mathrm{Re}[\Phi(\mathbf k)]\nonumber.
\end{eqnarray}
It should be noted that the above equations are simple reformulation of Eq. \ref{alm}.
From Eq. \ref{alm2} and \ref{alm3}, we may see that the odd-parity preference might be produced, provided
\begin{eqnarray}
|\mathrm{Re} [\Phi(\mathbf k)]|\ll|\mathrm{Im} [\Phi(\mathbf k)]|\;\;\;(k\lesssim 22/\eta_0),\label{primordial_odd} 
\end{eqnarray}
where $\eta_0$ is the present conformal time.
Taking into account the reality condition $\Phi(-\mathbf k)= \Phi^*(\mathbf k)$, we may show primordial perturbation in real space is given by:
\begin{eqnarray}
\Phi(\mathbf x)&=&2\int\limits^{\infty}_0 dk  \int\limits^{\pi}_0 d\theta_{\mathbf k} \sin\theta_{\mathbf k}\int\limits^{\pi}_0 d\phi_{\mathbf k}\label{Phi_real}\\
&\times&\left(\mathrm{Re}[\Phi(\mathbf k)]\cos(\mathbf k\cdot \mathbf x)-\mathrm{Im}[\Phi(\mathbf k)]\sin(\mathbf k\cdot \mathbf x)\right). \nonumber
\end{eqnarray}
Noting Eq. \ref{primordial_odd} and \ref{Phi_real}, we find our primordial Universe may possess odd-parity preference on large scales ($2/\eta_0 \lesssim k\lesssim 22/\eta_0$).
The odd-parity preference of our primordial Universe violates large-scale translational invariance in all directions.
However, it is not in direct conflict with the current data on observable Universe (i.e. WMAP CMB data), though it may seem intriguing. 
Considering Eq. \ref{primordial_odd} and \ref{Phi_real}, we find this effect will be manifested on the scales larger than $2\pi\,\eta_0/22\approx 4\,\mathrm{Gpc}$.
However, it will be difficult to observe such large-scale effects in non-CMB observations.
If the odd-parity preference is indeed cosmological, it indicates we are at a special place in the Universe, which may sound intriguing.
However, it should be noted that the invalidity of the Copernican Principle such as our living near the center of void had been previously proposed in different context \citep{Void_DE,Void_SN}.

Depending on the type of cosmological origins (e.g. topology, features in primordial power spectrum and Eq. \ref{primordial_odd}), distinct anomalies are predicted in polarization power spectrum. Therefore, polarization maps of large-sky coverage (i.e. low multipoles) will allow us to remove degeneracy and
figure a cosmological origin, if the parity asymmetry is indeed cosmological.

\section{Phase of even and odd multipole data}
\label{phase}
The decomposition coefficients $a_{lm}$ of CMB anisotropy, which is briefly discussed in Section \ref{asymmetry}, is equivalently written as:
\begin{eqnarray} 
a_{lm} = |a_{lm}| \exp(i \phi_{lm}). 
\end{eqnarray}
Given a Gaussian model, we expect that the amplitudes $|a_{l,m}|$ and the phase $\phi_{lm}$ follow the Rayleigh distribution and a uniform distribution $[0,2 \pi]$ respectively \citep{Bardeen,Inflation,Foundations_Cosmology}.
Therefore, the phase information provides additional information on the statistical properties and hence useful test on Gaussianity. 
Noting this, we have investigated the phases and compared those of even and odd multipole data.
For the analysis, we are going to use the following trigonometric moments:
\begin{eqnarray}
\label{si_org}
\mathcal S(l) &=& \frac{1}{2l +1} \sum^{l}_{m=-l} \sin(\varphi_{lm}),\\
\mathcal C(l) &=& \frac{1}{2l+1} \sum^{l}_{m=-l} \cos(\varphi_{lm})
\end{eqnarray}
Using the trigonometric moments, we may estimate the mean angle $\Theta(l)$ as follows:
\begin{eqnarray}\label{mean_angle}
\Theta(l) = \arctan\left(\frac{\mathcal S(l)}{\mathcal C(l)}\right),
\end{eqnarray}
where the information of the CMB phases is condensed into a single mean angle for an individual multipole.
Further details on the procedures above can be found in \citep{circular_data}.
Given a Gaussian random Universe, we would expect the mean angles of each multipoles to follow a uniform distribution ($-\pi\le \Theta\le \pi$) \citep{COBE_NG}.
In order to investigate the even and odd multipole data respectively, we will use the following statistics: 
\begin{eqnarray}\label{m_a}
r^{\pm}_s(l_{\mathrm{max}})&=&\sum_{l=2}^{l_{\mathrm{max}}}\sin\Theta^{\pm}(l),\\
r^{\pm}_c(l_{\mathrm{max}})&=&\sum_{l=3}^{l_{\mathrm{max}}}\cos\Theta^{\pm}(l)\,\\
R^{\pm}(l_{\mathrm{max}})&=&\frac{1}{l_{\mathrm{max}}-2}\left([r^{\pm}_s(l_{\mathrm{max}})]^2+[r^{\pm}_c(l_{\mathrm{max}})]^2\right),\nonumber\\
\end{eqnarray}
where we imply the quantities of $+$ and $-$ are associated with the even and odd multipole data respectively.
In the theory of statistical analysis of circular data, the statistic $R$ is widely used (refer to \citep{directional_statistics} for details).
After a simple algebra on Eq(\ref{m_a}), we may easily show:
\begin{eqnarray}\label{R_a}
R^{\pm}(l_{\mathrm{max}})&=&\frac{1}{l_{\mathrm{max}}-2}\sum_{ll'}^l\cos\left(\Theta^{\pm}(l)-\Theta^{\pm}(l')\right)\nonumber\\
&=&\frac{1}{2}\left(\delta_{l_{\mathrm{max}},2n}+\frac{l_{\mathrm{max}}-1}{l_{\mathrm{max}}-2}\delta_{l_{\mathrm{max}},2n+1}\right)\\
&& +\frac{1}{l_{\mathrm{max}}-2}\sum_{l,l'\neq l}^l\cos\left(\Theta^{\pm}(l)-\Theta^{\pm}(l')\right)\nonumber
\end{eqnarray}
If the mean angles are correlated among distinct multipoles, the second term in Eq. \ref{R_a} is given by $\sim l^{-2}$, and thus
$R^{\pm}$ asymptotically approaches $1/2$ \citep{Phase_NG}.
On the other hand, if the mean angles of distinct multipoles are similar to each other (i.e. $\cos\left(\Theta^{\pm}(l)-\Theta^{\pm}(l')\right)\sim \pm 1$), the second term in Eq(\ref{R_a}) is comparable to the first term, and thus $R^{\pm}$ asymptotically approaches $1$.

In order to investigate the correlation of mean angles, we have estimated $\cos(\Theta(l) - \Theta(l+\Delta l))$.
For $\Delta l=1$ and $\Delta l=2$, we find several unusual alignment for various multipoles, where 
the unusual alignment is found in Galactic coordinate and Ecliptic coordinate respectively for $\Delta l=1$ and $\Delta l=2$.
In Table \ref{EstimatorVal1} and \ref{EstimatorVal2}, we show the values of $\cos(\Theta(l) - \Theta(l+1))$ and $\cos(\Theta(l) - \Theta(l+2))$.
\begin{table}[htb!]
\centering
\caption{Mean angle correlation of WMAP data in Galactic coordinate}
\begin{tabular}{cc}
\hline 
$l$ &\;\;$\cos(\Theta(l) - \Theta(l+1))$ \\
\hline
$2$  & $0.9714$ \\ 
$18$  & $0.9947$\\ 
$28$  & $0.9978$\\ 
$33$ & $0.9477$\\ 
$36$  & $0.9299$\\ 
$38$  & $0.9485$\\
\hline
\end{tabular}
\label{EstimatorVal1}
\end{table}
\begin{table}[htb!]
\centering
\caption{Mean angle correlation of WMAP data in Ecliptic coordinate}
\begin{tabular}{cc}
\hline 
$l$ &\;\;$\cos(\Theta(l) - \Theta(l+2))$ \\
\hline
$5$  & $0.9986$ \\ 
$23$  & $0.9995$\\ 
$33$  & $0.9998$\\ 
$34$ & $0.9999$\\ 
\hline
\end{tabular}
\label{EstimatorVal2}
\end{table}

\begin{figure}[htb!]
\centering\includegraphics[width=0.475\textwidth]{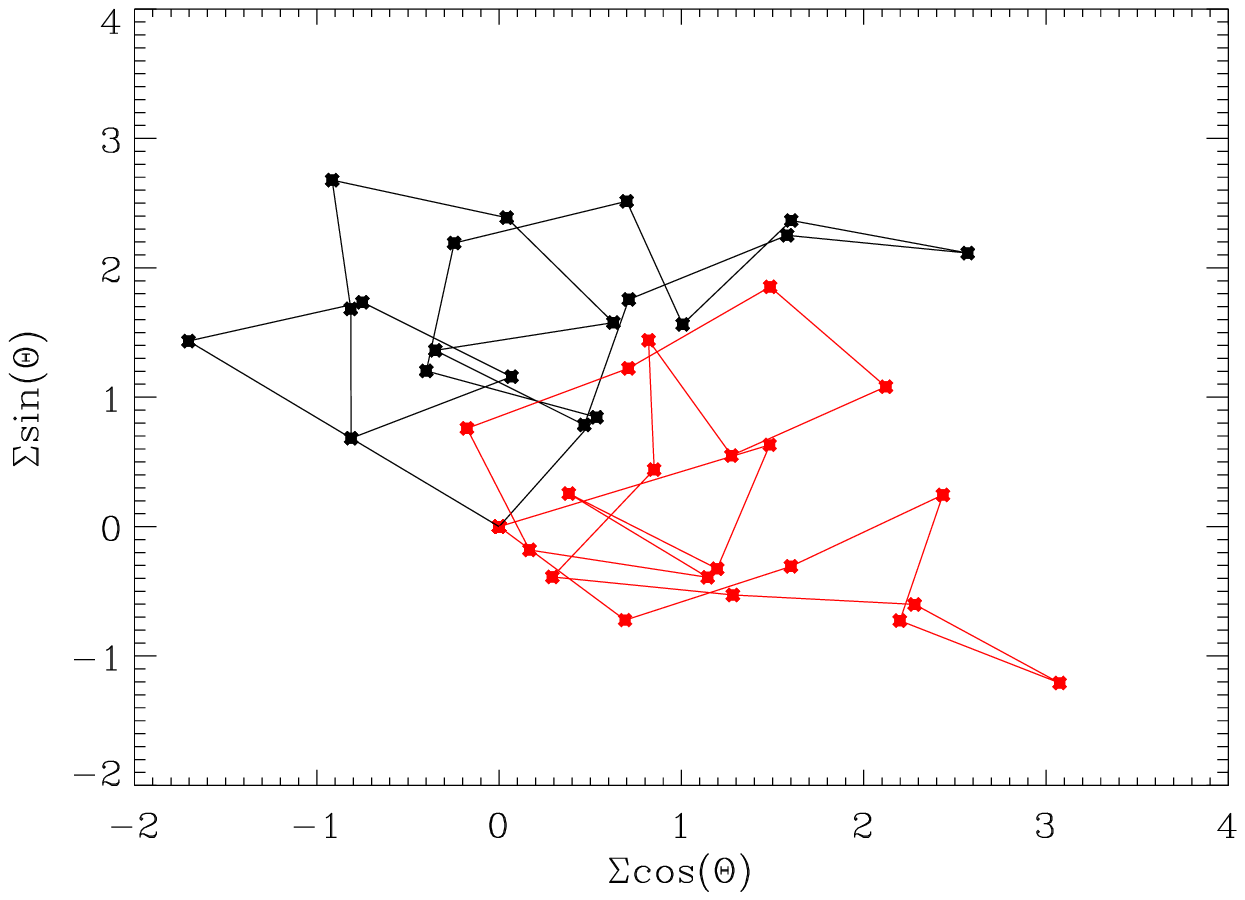}
\centering\includegraphics[width=0.475\textwidth]{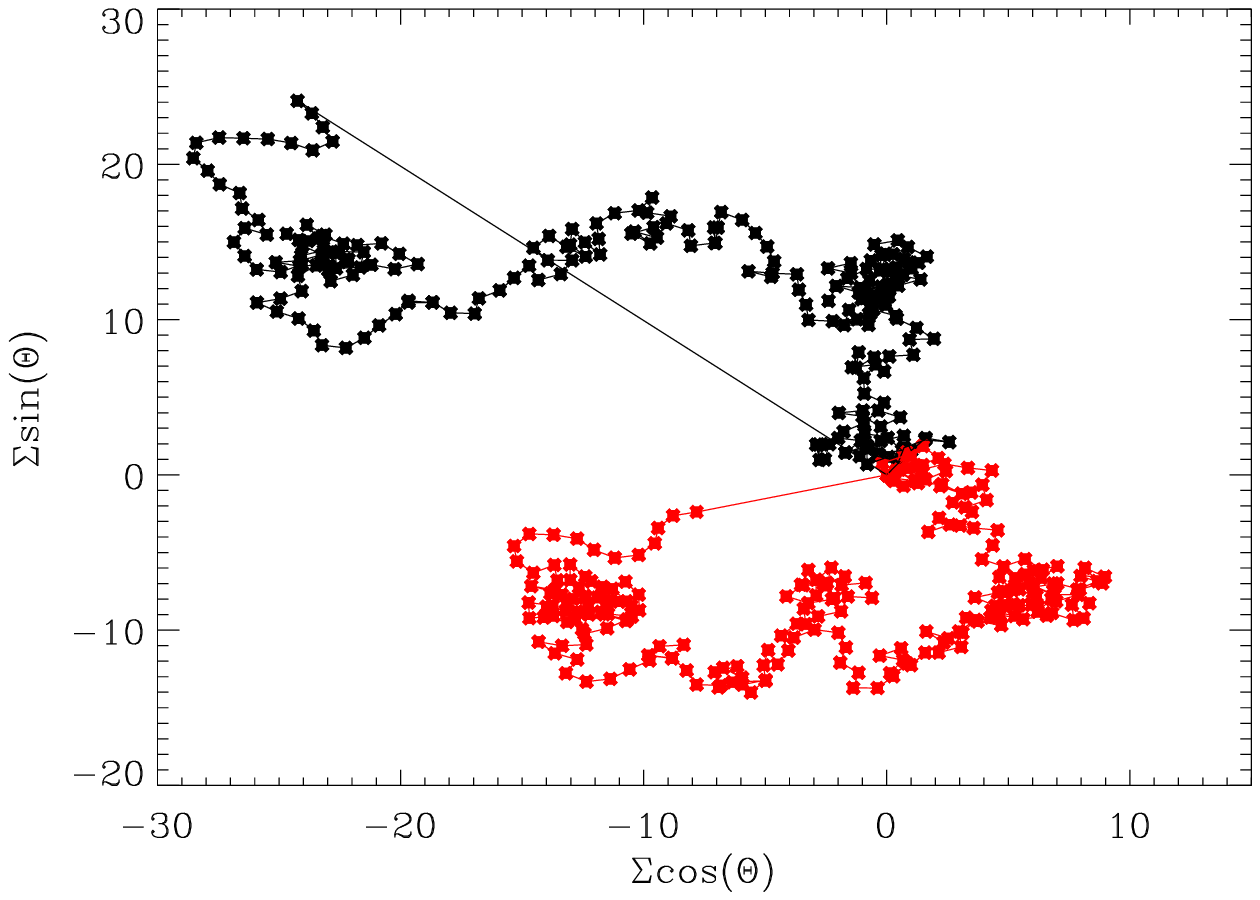}
\caption{Top panel. The parameter $r^-(l)$ (the black dots) for odd multipoles and $r^+(l)$ (the red dots) for even ones in Galactic coordinates. The mean angles is estimated from the first 40 multipoles. Bottom panels. The same as the top panels, but for ecliptic coordinates.}
\label{figure_mean_walk}
\end{figure}

\begin{figure}[htb!]
\centering\includegraphics[width=0.465\textwidth]{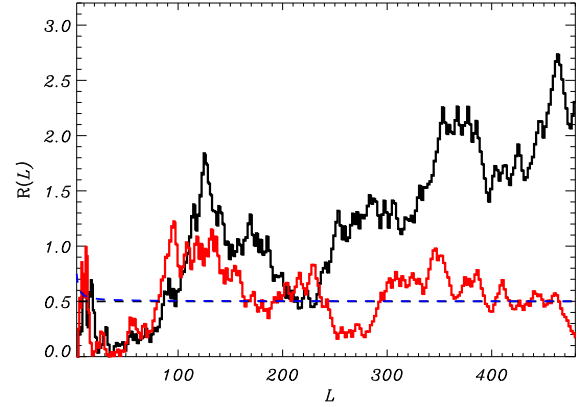}
\caption{The parameters $R^-(l)$ (top panel, the black line) for odd multipoles and $R^+(l)$ (top panel, the red line) for even multipoles. The black dash line is the asymptotic $R^+(l=even)=0.5$, the blue dash line is for $R^-(l=odd)=\frac{l-1}{2(l-2)}$.}
\label{figure_mean_w}
\end{figure}
In Fig. \ref{figure_mean_walk}, we plot $r^{\pm}(l)$ of the WMAP team's ILC map, which are estimated in Galactic coordinate and Ecliptic coordinate respectively.
As shown in the figure, the mean angles of the even and odd are distinct from each other, which is significant at $\sim 3\sigma$ level.
In Fig. \ref{figure_mean_w}, we show $R^{\pm}(l)$ and their inverse for various values of $l$, which are estimated in Galactic coordinate.
From this figure, we can see that a major contribution to the $R^-(l)$ comes from the multipole $5\le l\le 30$, where the 
$R^-(l)$ exceeds the $2\sigma$. The even multipoles, in contrast with the odd ones, show noticeably small values of the
$R^+(l)$-parameter, which indicate the correlations between the mean angles.
It should be noted that values of $R^-(l)$ and $R^+(l)$ are expected to be around $\sim 0.5$. 
Therefore, unusually high or small values as those of WMAP data indicates unusual correlation of mean angles, and the deviation from statistically isotropic Gaussian Universe.

\section{antisymmetry of the octupole component}
\label{S3}
Using Eq. \ref{T_expansion} and the reality condition $a_{lm}=a^*_{l-m}$, we may easily show that a whole-sky CMB anisotropy pattern is given by:
\begin{eqnarray}
T(\theta,\phi) &=&\sum_{l} a_{l0}\,N_{l0}\,P_l(\cos\theta)+2\sum_{l}\sum_{m\ge1} \,N_{lm}\,P^m_l(\cos\theta)\nonumber\\
&\times&(\mathrm{Re}[a_{lm}]\,\cos(m\phi)-\mathrm{Im}[a_{lm}]\,\sin(m\phi)),\label{T_expansion2}
\end{eqnarray}
where
$P_l(\cos\theta)$ and $P^m_l(\cos\theta)$ are the Legendre polynomials and the associated Legendre polynomials respectively, and
\begin{eqnarray}
N_{lm}=(-1)^m\sqrt{\frac{(2l+1)(l-m)}{4\pi(l+m)!}}.
\end{eqnarray}

\begin{figure}[htb!]
\centering\includegraphics[width=0.32\textwidth]{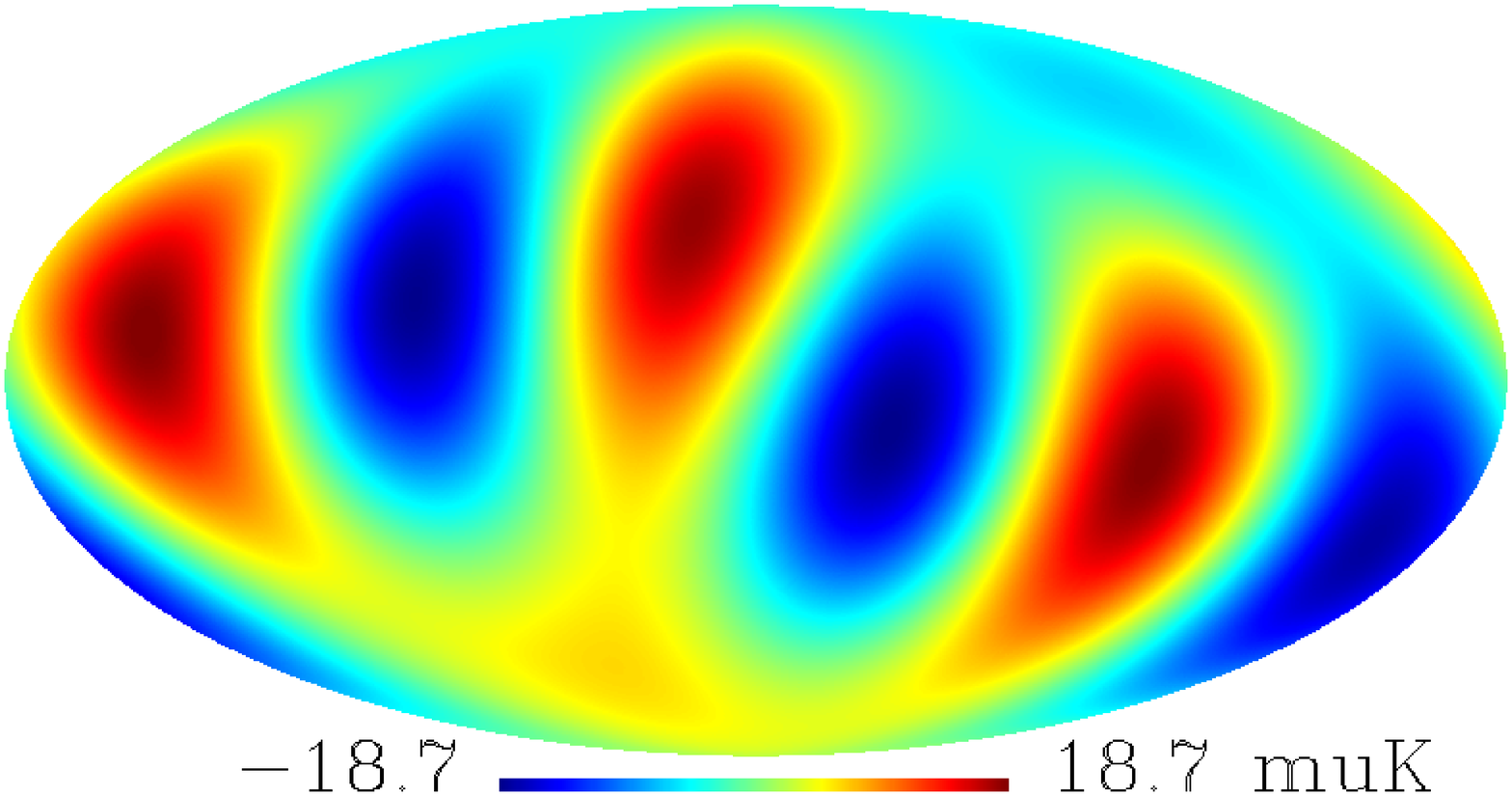}
\centering\includegraphics[width=0.32\textwidth]{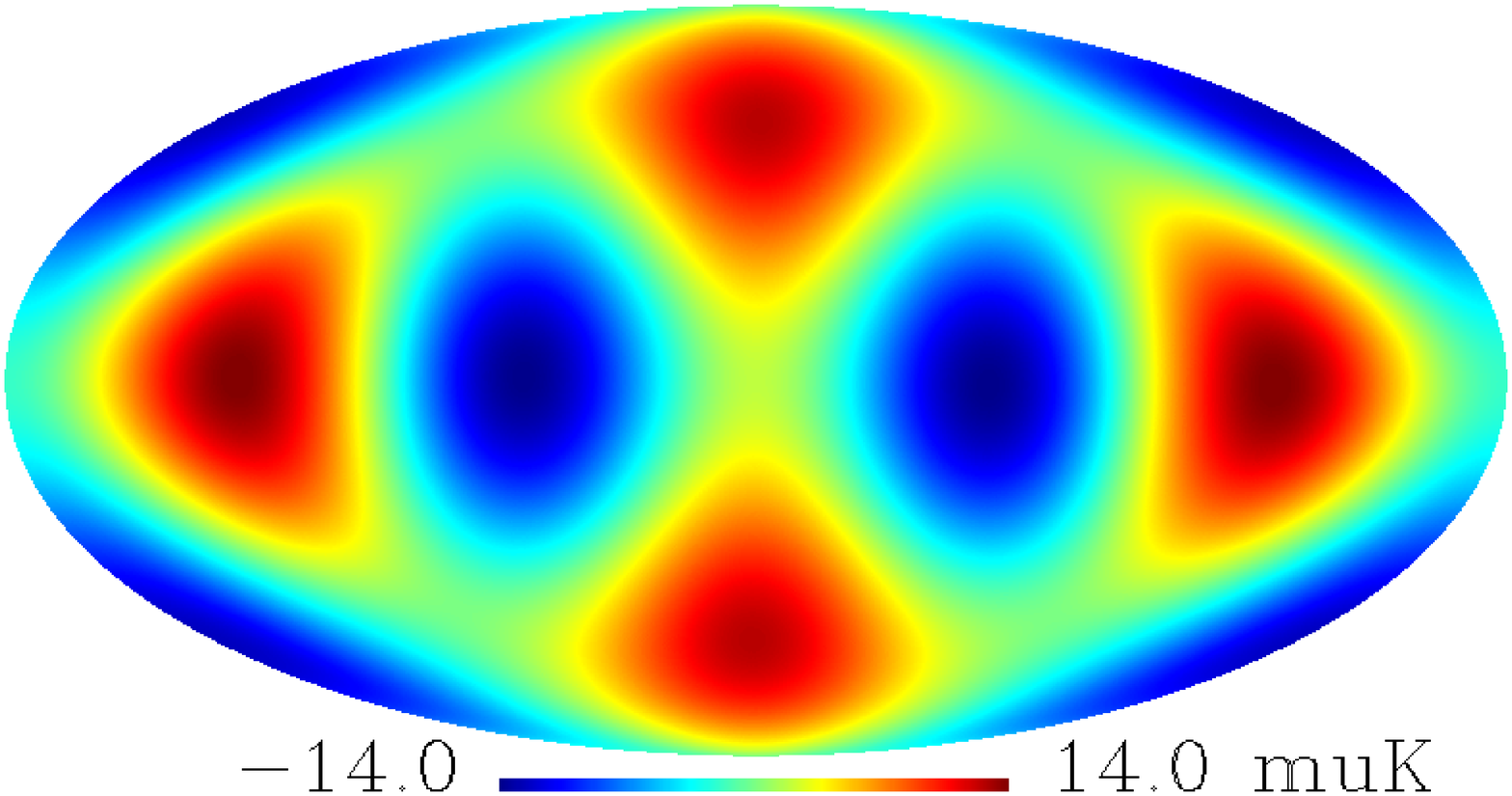}
\centering\includegraphics[width=0.32\textwidth]{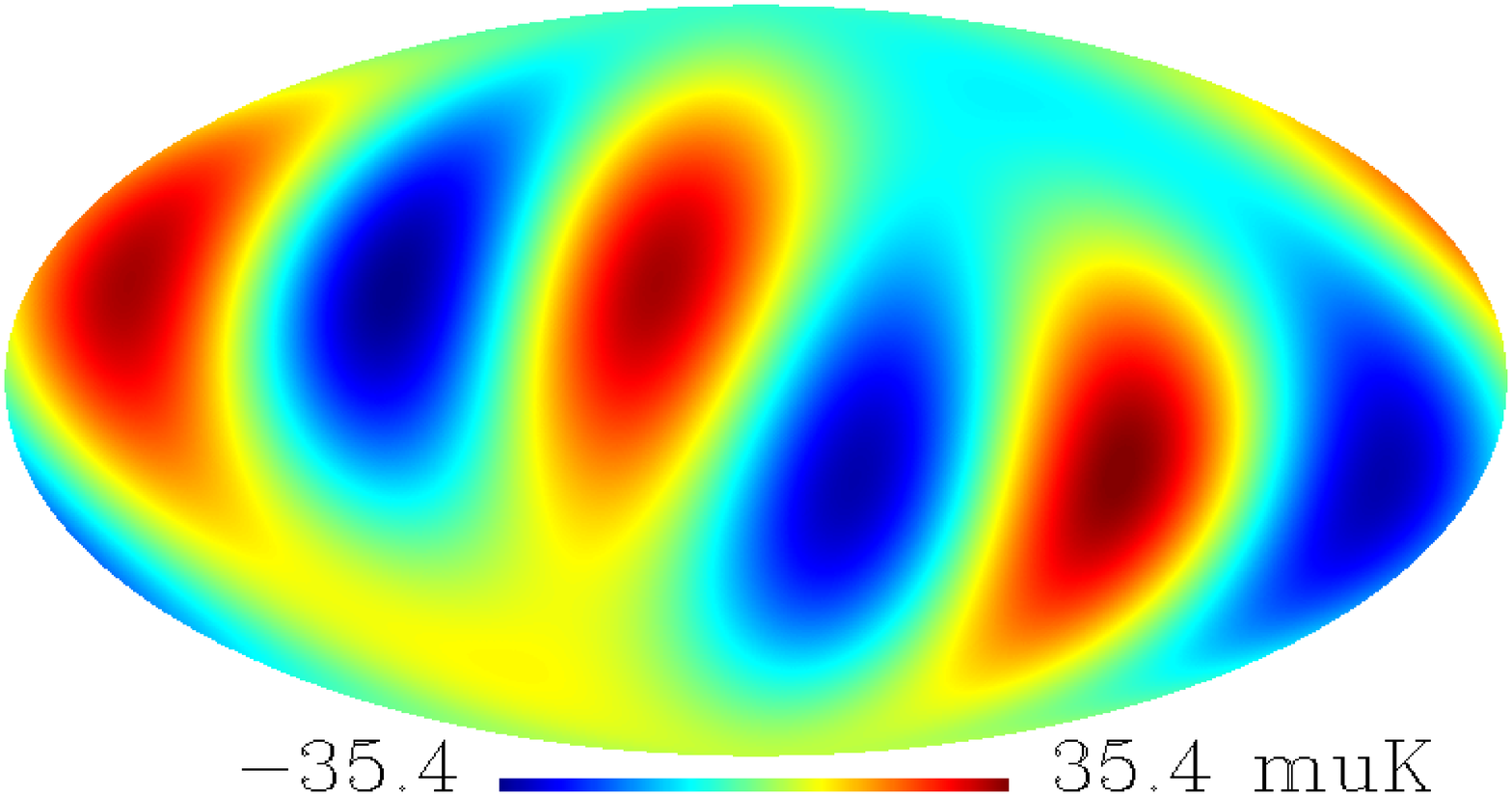}
\caption{The octupole components of the WMAP7 ILC map (top), $T^{+}_3(\theta,\phi)$ (middle), and $T^{-}_3(\theta,\phi)$ (bottom)}
\label{octupole}
\end{figure}
In addition to the parity operation (i.e. $\mathbf n\rightarrow -\mathbf n$) discussed previously, we may consider the following coordinate inversion:
$(\theta,\phi)\rightarrow (\pi-\theta,-\phi)$.
In a similar way to the investigation on the parity asymmetry, we construct a symmetric and antisymmetric part under the coordinate inversion $(\theta,\phi)\rightarrow (\pi-\theta,2\pi-\phi)$ as follows:
\begin{eqnarray} 
T^{\pm}(\theta,\phi)&=&\frac{T(\theta,\phi)\pm T(\pi-\theta,-\phi)}{2},
\end{eqnarray}
Using Eq. \ref{T_expansion2}, we may easily show the symmetric and antisymmetric parts are given by:
\begin{eqnarray}
\lefteqn{T^{\pm}(\theta,\phi)=\sum_{l}\,N_{l0}\,P_l(\cos\theta)\,\mathrm{Re}[a_{l0}]\,\frac{1\pm(-1)^l}{2}}\nonumber\\
&+&2\sum_{l}\sum_{m\ge1}N_{lm}P^m_l(\cos\theta)\left(\mathrm{Re}[a_{lm}]\frac{1\pm(-1)^{l+m}}{2}\right.\nonumber\\
&&\times\left.\cos(m\phi)-\mathrm{Im}[a_{lm}]\frac{1\mp(-1)^{l+m}}{2}\,\sin(m\phi)\right).\label{T_pm_expansion}
\end{eqnarray}
As obvious in Eq. \ref{T_pm_expansion}, the symmetric part gets contribution only from $\mathrm{Re}[a_{lm}]$ of $l+m=$even and $\mathrm{Im}[a_{lm}]$ of $l+m=$odd.
On the other hand, the antisymmetric part gets contribution only from $\mathrm{Re}[a_{lm}]$ of $l+m=$odd and $\mathrm{Im}[a_{lm}]$ of $l+m=$even.
Noting this, we have estimated the ratio of $l+m=$odd and  $l+m=$even components for the real and imaginary parts, where we used the WMAP 7 year ILC map.
Our estimation shows the ratio $\mathrm{Im}[a_{33}]/\mathrm{Im}[a_{32}]$ is unusually high, which requires the chance of 6-in-1000 level.
In Fig. \ref{octupole}, we show the octupole components of the WMAP7 ILC map (top), $T^+$ (middle) and $T^-$ (bottom).
From Fig. \ref{octupole}, we may see the octupole components of the WMAP7 ILC map shows antisymmetric pattern for the inversion $(\theta,\phi)\rightarrow (\pi-\theta,-\phi)$, where the center of the images corresponds to the coordinate $(\theta=0,\phi=0)$.

\section{Parametric tension between even and odd multipole data}
\label{cosmomc}
As discussed previously, there is the power contrast between even and odd multipoles of WMAP $TT$ power spectrum \citep{Universe_odd,odd,odd_origin,odd_bolpol,WMAP7:anomaly}.
At lowest multipoles ($2\le l\le 22$), there is odd multipole preference (i.e. power excess in odd multipoles and deficit in even multipoles) \citep{Universe_odd,odd,odd_origin,odd_bolpol}, and additionally even multipole preference at intermediate multipoles ($200\le l \le400$) \citep{WMAP7:anomaly}.
For TE correlation, we have also found odd multipole preference at ($100\lesssim l\lesssim 200$) and even multipole preference at ($200\lesssim l \lesssim400$), though its statistical significance is not high enough, due to low Signal-to-Noise Ratio of polarization data.
Not surprisingly, these power contrast anomalies are explicitly associated with the angular power spectrum data, which are mainly used to fit cosmological models. 
Having noted this, we have investigated whether the even(odd) multipole data set is consistent with the concordance model.
\begin{figure}[htb!]
\centering
\includegraphics[scale=.55]{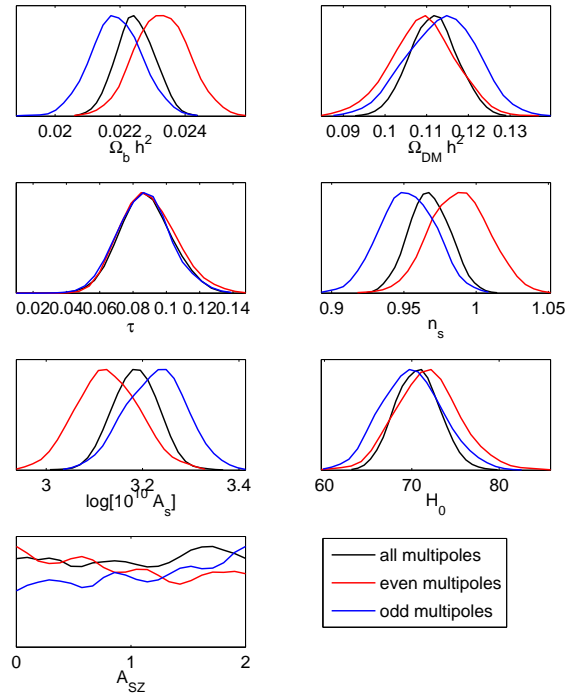}
\caption{Marginalized likelihood of cosmological parameters ($\Lambda$CDM + sz + lens), given whole or even(odd) multipole data.}
\label{like1}
\end{figure}
For a cosmological model, we have considered $\Lambda$CDM + SZ effect + weak-lensing, where cosmological parameters are $\lambda \in \left\{\Omega_b,\Omega_{c},\tau,n_s, A_s, A_{sz}, H_0 \right\}$. For data constraints, we have used the WMAP 7 year TT  and TE power spectrum data, which have been estimated from the ILC map, and cut-sky V and W band maps \citep{WMAP7:powerspectra}.
Hereafter, we shall denote WMAP CMB data of whole, even and odd multipoles by $D_0$, $D_2$ and $D_3$ respectively.
We like to stress  that even/odd multipole splitting are made for TT and TE power spectrum up to the multipoles of WMAP sensitivity (i.e. $l\le 1200$ for TT and $l\le 800$ for TE). Using \texttt{CosmoMC} with the modified WMAP likelihood code, we have explored the parameter space on a MPI  cluster with 6 chains \citep{CosmoMC,Gibbs_power,CosmoMC_note,WMAP7:powerspectra}.
For the convergence criterion, we have adopted the Gelman and Rubin's ``variance of chain means'' and set the R-1 statistic to $0.03$ for stopping criterion \citep{Gelman:inference,Gelman:R1}.

In Fig. \ref{like1}, we show the marginalized likelihood of parameters, which are obtained from the run of a \texttt{CosmoMC} with $D_0$, $D_2$ and $D_3$ respectively.
\begin{table}[htb!]
\centering
\caption{cosmological parameters ($\Lambda$CDM + sz + lens)}
\begin{tabular}{cccc}
\hline\hline 
 & $\lambda_0$  &$\lambda_2$   & $\lambda_3$ \\
\hline
$\Omega_{b}\,h^2$  & $0.0226\pm 0.0006$ &$0.0231\pm0.0008$ & $0.0217\pm0.0008$ \\ 
$\Omega_{c}\,h^2$  & $0.112\pm0.006$ &$0.109\pm0.008$ & $0.115\pm0.008$ \\ 
$\tau$  & $0.0837\pm 0.0147$ &$0.0913\pm0.0157$ & $0.0859\pm0.015$ \\ 
$n_s$ & $0.964\pm 0.014$ &$0.989\pm0.02$ & $0.949\pm0.019$ \\ 
$\log[10^{10} A_s]$  & $3.185\pm0.047$ &$3.132\pm0.065$ & $3.239\pm0.062$ \\ 
$H_0$  & $70.53\pm2.48$ &$71.73\pm3.59$ & $69.68\pm3.47$ \\
$A_{\mathrm{sz}}$  & $1.891^{+0.109}_{-1.891}$ &$0.169^{+1.831}_{-0.169}$ & $0.89^{+1.11}_{-0.89}$ \\ 
\hline
\end{tabular}
\label{parameter1}
\end{table}
In Table \ref{parameter1}, we show the best-fit parameters and 1 $\sigma$ confidence intervals, where $\lambda_2$ and $\lambda_3$ denote the best-fit values of $D_2$ and $D_3$ respectively. The parameter set $\lambda_0$ are the best-fit values of whole data $D_0$, and accordingly corresponds to the WMAP concordance model.
As shown in Fig. \ref{like1} and Table \ref{parameter1}, we find non-negligible tension especially in parameters of primordial power spectrum.
It is worth to note that the best-fit spectral index of even multipole data (i.e. $D_2$) is close to a flat spectrum (i.e. $n_s=1$),  while the result from the whole data rule out the flat spectrum by more than 2$\sigma$.

There is a likelihood-ratio test, which allows us to determine the rejection region of an alternative hypothesis, given a null hypothesis \citep{statistics_theory,theoretical_statistics,Statistics_Lupton,Math_methods}.
By setting sets of parameters to a null hypothesis and an alternative hypothesis, we may investigate whether two sets of parameters are consistent with each other.
To be specific, we have evaluated the following in order to assess parametric tension:
\begin{eqnarray*} 
\frac{\mathcal L(\lambda_j|D_i)}{\mathcal L(\lambda_i|D_i)},
\end{eqnarray*}
where parameter set $\lambda_i$ and $\lambda_j$ correspond to a null hypothesis and an alternative hypothesis respectively. 
\begin{table}[htb!]
\centering
\caption{the likelihood ratio: $\Lambda$CDM + sz + lens}
\begin{tabular}{c|ccc}
\hline\hline
 & $\mathcal L(\lambda_0|D_0)$  & $\mathcal L(\lambda_2|D_0)$    & $\mathcal L(\lambda_3|D_0)$  \\
\hline $\mathcal L(\lambda_0|D_0)$  & $1$ &$ 0.076$ & $0.0099$\\
\hline\hline 
& $\mathcal L(\lambda_0|D_2)$  & $\mathcal L(\lambda_2|D_2)$    & $\mathcal L(\lambda_3|D_2)$  \\
\hline
$\mathcal L(\lambda_2|D_2)$  & $0.16$ &$1$ & $2\times 10^{-4}$ \\ 
\hline\hline
& $\mathcal L(\lambda_0|D_3)$  & $\mathcal L(\lambda_2|D_3)$    & $\mathcal L(\lambda_3|D_3)$  \\
\hline
$\mathcal L(\lambda_3|D_3)$   & $0.16$ &$0.0022$ & $1$ \\ 
\hline
\end{tabular}
\label{fit1}
\end{table}
In Table \ref{fit1}, we show the likelihood ratio, where the quantities used for the numerator and denominator are indicated in the uppermost row and leftmost column.
As shown by $\mathcal L(\lambda_0|D_2)/\mathcal L(\lambda_2|D_2)$ and $\mathcal L(\lambda_0|D_3)/\mathcal L(\lambda_3|D_3)$, the WMAP concordance model (i.e. $\lambda_0$) does not make a good fit for even(odd) multipole data set.
Besides, there exist significant tension between two data subsets, as indicated by very small values of 
$\mathcal L(\lambda_3|D_2)/\mathcal L(\lambda_2|D_2)$ and $\mathcal L(\lambda_2|D_3)/\mathcal L(\lambda_3|D_3)$.
The parameter likelihood, except for $A_{\mathrm{sz}}$, follows the shape of Gaussian functions, as shown in Fig. \ref{like1}.
For a likelihood of Gaussian shape, the likelihood ratio 0.1353 and 0.0111 correspond to 2$\sigma$ and 3$\sigma$ significance level respectively.
From Table \ref{fit1}, we may see most of the ratio indicates $\sim 2\sigma$ tension or even higher. 

\begin{figure}[htb!]
\centering
\includegraphics[scale=.55]{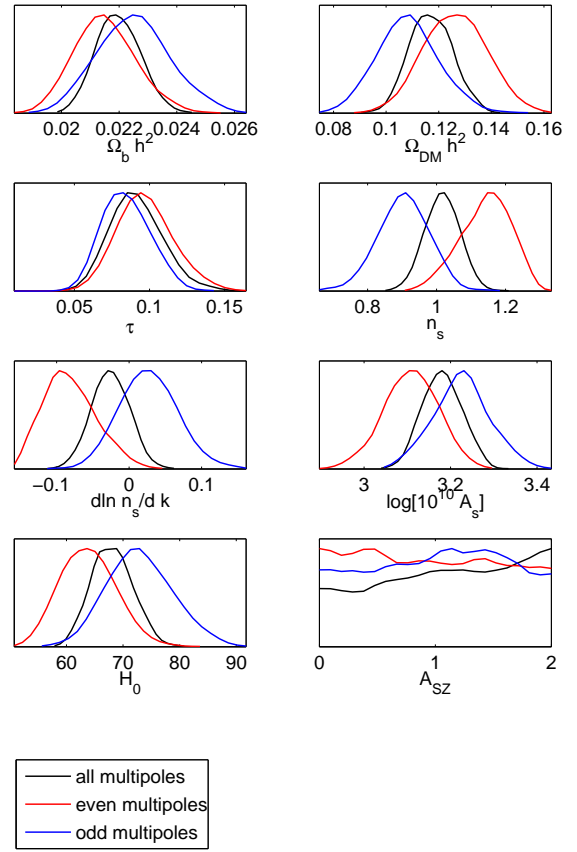}
\caption{Marginalized likelihood of cosmological parameters ($\Lambda$CDM + sz + lens + run), given whole, even and odd multipole data respectively.}
\label{like2}
\end{figure}

\begin{table}[htb!]
\centering
\caption{the likelihood ratio: $\Lambda$CDM + sz + lens + run}
\begin{tabular}{c|ccc}
\hline\hline
 & $\mathcal L(\lambda_0|D_0)$  & $\mathcal L(\lambda_2|D_0)$    & $\mathcal L(\lambda_3|D_0)$  \\
\hline $\mathcal L(\lambda_0|D_0)$  & $1$ & $3.5\times 10^{-4}$ & $0.0078$\\
\hline\hline 
& $\mathcal L(\lambda_0|D_2)$  & $\mathcal L(\lambda_2|D_2)$    & $\mathcal L(\lambda_3|D_2)$  \\
\hline
$\mathcal L(\lambda_2|D_2)$  & $0.06$ &$1$ & $2.3\times 10^{-5}$ \\ 
\hline\hline
& $\mathcal L(\lambda_0|D_3)$  & $\mathcal L(\lambda_2|D_3)$    & $\mathcal L(\lambda_3|D_3)$  \\
\hline
$\mathcal L(\lambda_3|D_3)$   & $0.042$ &$5.8\times 10^{-7}$ & $1$ \\ 
\hline
\end{tabular}
\label{fit2}
\end{table}

As discussed previously, the tension is highest in parameters of primordial power spectrum, which may be an indication of missing parameters in primordial power spectrum (e.g. a running spectral index). Therefore, we have additionally considered a running spectral index $dn_s/d \ln k$ and repeated our investigation.
Surprisingly, we find tension increases to even a higher level. 
We show the marginalized parameter likelihoods and the likelihood ratios in Fig. \ref{like2} and Table \ref{fit2}, where we find tension is also highest in the primordial power spectrum parameters. 
These tension indicate there is either unaccounted contamination or the failure of the assumed cosmological model (i.e. the flat $\Lambda$CDM model).

\section{Lack of angular correlation in the WMAP data}
\label{correlation}
Given CMB anisotropy data, we may estimate two point angular correlation as follows:
\begin{eqnarray}
C(\theta)= T(\hat{\mathbf n}_1)\:T(\hat{\mathbf n}_2),
\end{eqnarray}
where $\theta=\cos^{-1}(\hat{\mathbf n}_1\cdot \hat{\mathbf n}_2)$.
Using Eq. \ref{T_expansion} and \ref{alm_cor}, we may easily show that the expectation value of the correlation is given by \citep{structure_formation}:
\begin{eqnarray}
\langle C(\theta) \rangle=\sum_l \frac{2l+1}{4\pi}\,W_l\,C_l\,P_l(\cos\theta), \label{cor}
\end{eqnarray}
where $\theta$ is a separation angle, $W_l$ is the window function of the observation and $P_l$ is a Legendre polynomial. 
As shown in Eq. \ref{cor}, the angular correlation $C(\theta)$ is the linear combination of angular power spectrum $C_l$, and therefore, possess equivalence.

\begin{figure}[htb!]
\centering\includegraphics[scale=.55]{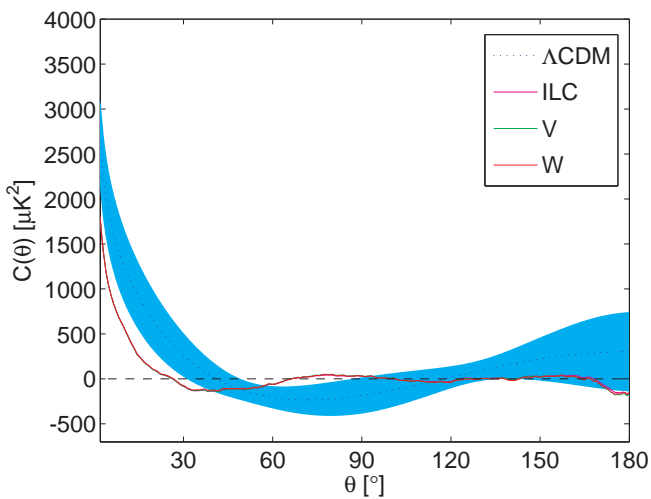}
\caption{Angular correlation of CMB anisotropy: Solid lines denote the angular correlation of WMAP data. Dotted line and shaded region denote the theoretical prediction and 1$\sigma$ ranges, as determined by Monte-Carlo simulations ($\Lambda$CDM).}
\label{C_data}
\end{figure}

In Fig. \ref{C_data}, we show the angular correlation of the WMAP 7 year data, which are estimated respectively from the WMAP team's Internal Linear Combination (ILC) map, and foreground reduced maps of V and W band. 
In the angular correlation estimation, we have excluded the foreground-contaminated region by applying the WMAP KQ75 mask, as recommended for non-Gaussianity study \citep{WMAP7:fg}.
In the same plot, we show the angular correlation of the WMAP concordance model \citep{WMAP7:Cosmology}, where the dotted line and shaded region denote the 
mean value and 1$\sigma$ ranges of Monte-Carlo simulations at V band.
For simulation, we have made $10^4$ realizations with the same configuration with the WMAP data (e.g. a foreground mask, beam smoothing and instrument noise).
In order to include WMAP noise in simulation, we have subtracted one Differencing Assembly (D/A) data from another, and added it to simulations.

As shown in Fig. \ref{C_data}, there exists non-negligible discrepancy between the data and the theoretical prediction.
Most noticeably, angular correlation of WMAP data nearly vanishes at angles larger than $\sim 60^\circ$, which are previously investigated by 
\citep{correlation_COBE,WMAP1:Cosmology,correlation_Copi1,correlation_Copi2,lowl_anomalies}.
In the previous investigations, the lack of large-angle correlation has been assessed by the following statistic \citep{WMAP1:Cosmology,correlation_Copi1,correlation_Copi2,lowl_anomalies}:
\begin{eqnarray}
S_{1/2}=\int^{1/2}_{-1} \left( C(\theta) \right)^2 d(\cos\theta). \label{S12}
\end{eqnarray}
The investigation shows the $S_{1/2}$ estimated from WMAP data is anomalously low, which requires the chance $\lesssim 10^{-3}$ \citep{WMAP1:Cosmology,correlation_Copi1,correlation_Copi2,lowl_anomalies,lowl_bias}.
Besides the lack of correlation at large angles, we may see from Fig. \ref{C_data} that correlation at small angles tends to be smaller than the theoretical prediction.
Noting this, we have investigated the small-angle correlation with the following statistics:
\begin{eqnarray}
S_{\sqrt{3}/2}&=&\int^{1}_{\sqrt{3}/2} \left( C(\theta) \right)^2 d(\cos\theta),\label{S32}
\end{eqnarray}
where the square of the correlation is integrated over small angles ($0\le \theta\le30^\circ$).
Therefore, the values of $S_{\sqrt{3}/2}$ and $S_{1/2}$ corresponds to the integrated power at small and large angles respectively.
\begin{table}[htb!]
\centering
\caption{$S$ statistics of WMAP 7 year data}
\begin{tabular}{ccccc}
\hline\hline 
  & band &  angles&value [$\mu\mathrm{K}^4$]& $p$-value\\
\hline
$S_{1/2}$ & V  &  $60^\circ\le \theta\le180^\circ$&$1.42\times 10^{3}$  & $8\times 10^{-4}$\\
$S_{1/2}$ & W  &  $60^\circ\le \theta\le180^\circ$&$1.32\times 10^{3}$  & $6\times 10^{-4}$\\
\hline
$S_{\sqrt{3}/2}$ & V  & $0^\circ\le \theta\le30^\circ$ &$2.02\times 10^{4}$  & $3.2\times 10^{-3}$\\
$S_{\sqrt{3}/2}$ & W  & $0^\circ\le \theta\le30^\circ$ &$2.03\times 10^{4}$  & $3.2\times 10^{-3}$\\
\hline
\end{tabular}
\label{pvalue_S}
\end{table}
In Table \ref{pvalue_S}, we show $S_{1/2}$ and $S_{\sqrt{3}/2}$ of the WMAP 7 year data.
Recall that the slight difference between V and W band is due to the distinct beam size, and simulations are made accordingly for each band.
In the same table, we show the $p$-value, where the $p$-value denotes fractions of simulations as low as those of WMAP data.
As shown in Table \ref{pvalue_S}, WMAP data have unusually low values of $S_{1/2}$ and $S_{\sqrt{3}/2}$, as indicated by their $p$-value.
It is worth to note that the $p$-value of $S_{\sqrt{3}/2}$ corresponds to very high statistical significance, even though it is not as low as that of $S_{1/2}$.
In summary, we find anomalous lack of correlation at small angles in addition to large angles.

\begin{figure}[htb!]
\centering\includegraphics[scale=.44]{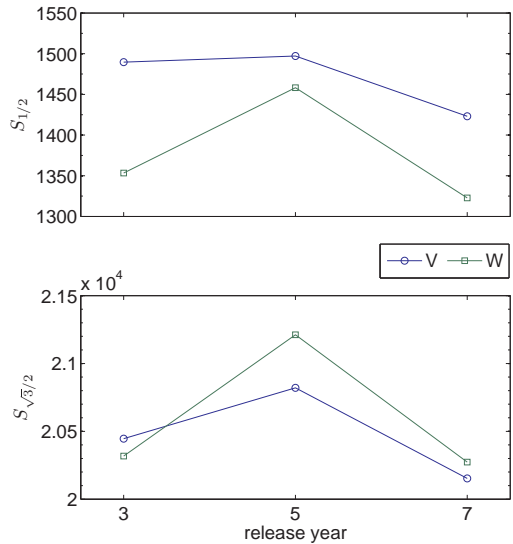}
\caption{$S$ statistics of WMAP 3, 5 and 7 year data}
\label{S_data}
\end{figure}
In Fig. \ref{S_data}, we show $S_{1/2}$ and $S_{\sqrt{3}/2}$, which are estimated from the WMAP 3, 5 and 7 year data respectively.
As shown in Fig. \ref{S_data}, the $S$ statistics of WMAP 7 year data are lowest, while WMAP 7 year data are believed to have more accurate calibration and less foreground contamination than earlier releases \citep{WMAP5:basic_result,WMAP7:basic_result,WMAP7:fg}. 
Therefore, we may not readily attribute the anomaly to calibration error or foregrounds. 

\subsection{Investigation on non-cosmological origins}
The WMAP data contain contamination from residual galactic and extragalactic foregrounds, even though 
we have applied the conservative KQ75 mask\citep{WMAP7:fg}.
\begin{figure}[htb!]
\centering\includegraphics[width=0.36\textwidth]{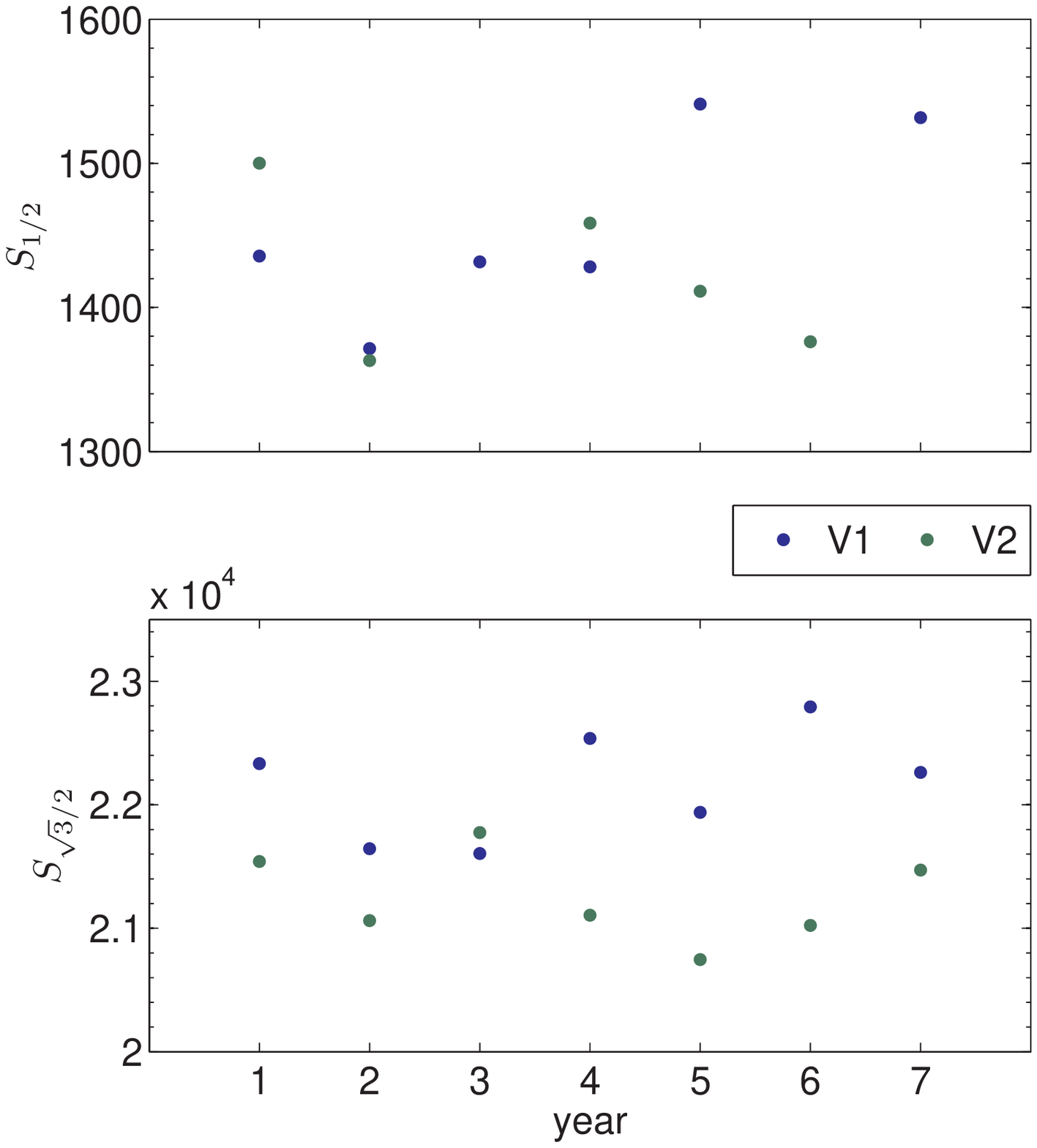}
\centering\includegraphics[width=0.36\textwidth]{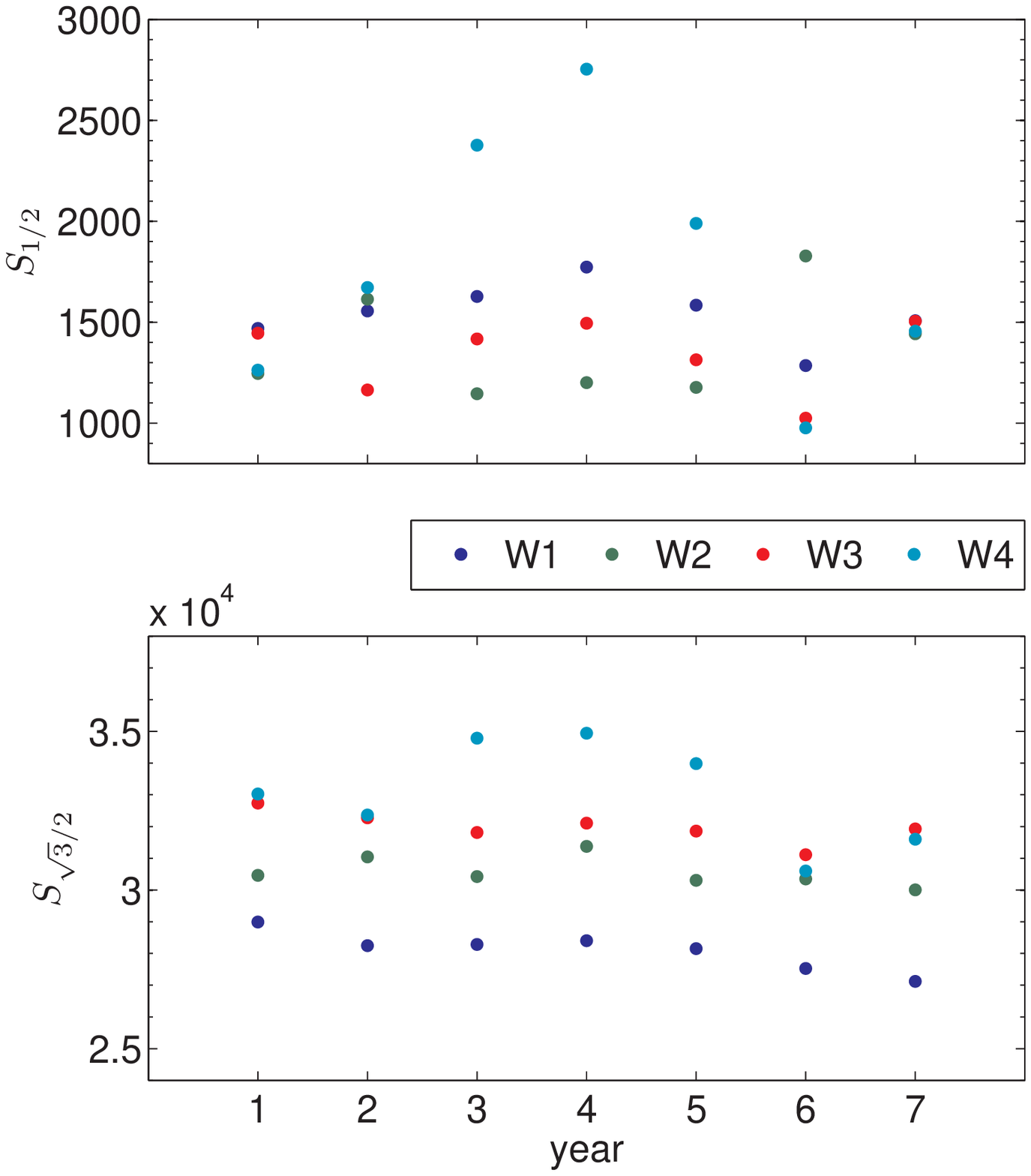}
\caption{the $S$ statistics of WMAP data at each D/A and year}
\label{S_year}
\end{figure}
\begin{figure}[htb!]
\centering\includegraphics[width=0.38\textwidth]{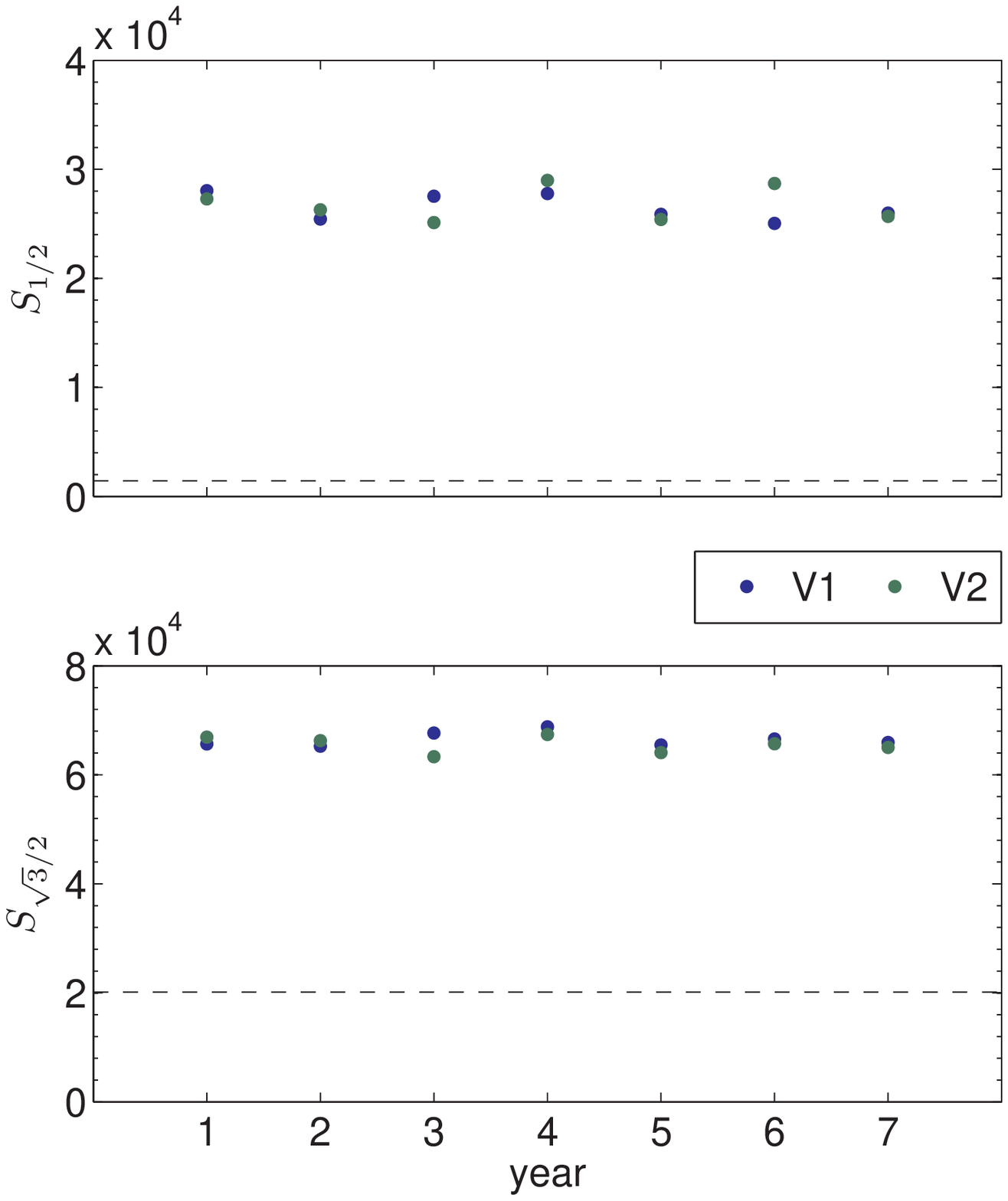}
\centering\includegraphics[width=0.38\textwidth]{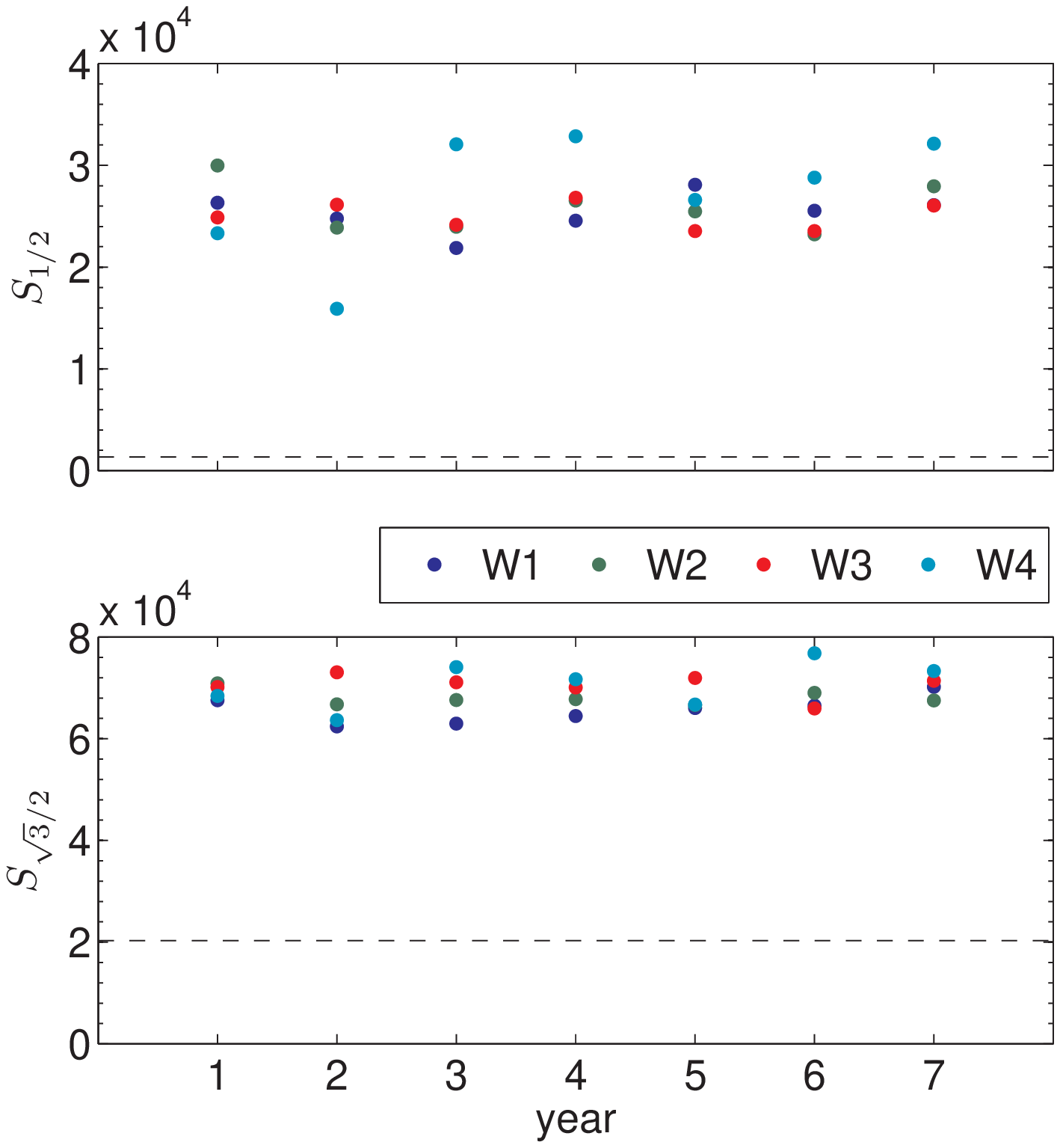}
\caption{the $S$ statistics of the simulated data produced by the WMAP team. Dashed lines show the values of WMAP data}
\label{S_sim}
\end{figure}
In order to investigate the association with residual foregrounds, we have first subtracted the foreground-reduced W band map from that of V band. 
This difference map mainly contains residual foregrounds of the forementioned maps with slight amount of CMB. Recall that CMB signal is not completely cancelled out, because the beam size at V and W band differs from each other.
From the difference map $V(\mathbf n)-W(\mathbf n)$, we have obtained $S_{1/2}=0.31$ and $S_{\sqrt{3}/2}=31.36$. By comparing these values with Table \ref{pvalue_S}, we may see residual foregrounds at V and W band are too small to affect the correlation power of WMAP data.
In order to investigate the association of noise with the anomaly, we have produced noise maps of WMAP7 data by subtracting one Differencing Assembly (D/A) map from another D/A data of the same frequency channel. 
\begin{table}[htb]
\centering
\caption{the $S$ statistics of WMAP instrument noise in the unit of [$\mu\mathrm{K}^4$]}
\begin{tabular}{ccc}
\hline\hline 
data &$S_{1/2}$ & $S_{\sqrt{3}/2}$ \\
\hline
V1-V2 & 0.25 &  83.94 \\
W1-W2 &  2.49 & 587.45 \\
W1-W3 &  2.18 & 664.26 \\
W1-W4 & 2.24 & 625.27 \\
W2-W3 & 2.72 & 808.32 \\
W2-W4 &  4.39 & 764.96\\
W3-W4 & 4.39 & 764.96\\
\hline
\end{tabular}
\label{S_noise}
\end{table}
In Table \ref{S_noise}, we show $S_{1/2}$ and $S_{\sqrt{3}/2}$ estimated from the noise maps.
Comparing Table \ref{pvalue_S} with Table \ref{S_noise}, we may see the noise is not significant enough to cause the correlation anomalies of the the WMAP data.
In Fig. \ref{S_year}, we show the values of $S_{1/2}$ and $S_{\sqrt{3}/2}$ for each year and D/A data. As shown in Fig. \ref{S_year}, the anomaly is not associated with a particular D/A channel nor a year data, but present at all year and D/A channels, which indicates the correlation anomaly is not due to the temporal malfunctioning of a particular D/A instrument. We have also investigated simulations produced by the WMAP team, which are discussed in the Section \ref{asymmetry}.
From the simulated maps, we have estimated $S_{1/2}$ and $S_{\sqrt{3}/2}$, which are plotted in Fig. \ref{S_sim}.
As shown in Fig. \ref{S_sim}, $S$ statistics of simulated data are significantly higher than those of WMAP data.
Therefore, the anomaly may be indeed cosmological or due to systematics, which we do not understand well.

\section{the parity asymmetry and the lack of correlation}
\label{association}
As shown in Eq. \ref{cor}, angular power spectrum and angular correlation possess some equivalence.
Noting this, we have investigated the association of the odd-parity preference with the lack of large-angle correlation, and found the odd-parity preference of the power spectrum is phenomenologically connected with the lack of large-angle correlation.
Using Eq. \ref{cor} with the Sach plateau approximation (i.e $l(l+1)\,C_l/2\pi\sim \mathrm{const}$), we find the expectation value of angular correlation is given by:
\begin{eqnarray}
C(\theta)&=&\sum_l \frac{2l+1}{4\pi}\,W_l\,C_l\,P_l(\cos\theta)\label{C_lowl}\\
&=&\sum_l \frac{l(l+1)\,C_l}{2\pi} \,\frac{2l+1}{2l(l+1)}\,W_l\,\,P_l(\cos\theta)\nonumber\\
&\approx&  \alpha \sum^{l_0}_{l} \frac{2l+1}{2l(l+1)}\,W_l\,\,P_l(\cos\theta)\\
&&+\sum_{l=l_0+1} C_l\,\frac{2l+1}{4\pi}\,W_l\,P_l(\cos\theta),\nonumber
\end{eqnarray}
where $\alpha$ is some positive constant and $l_0$ is a low multipole number, within which the Sach plateau approximation is valid.
As discussed previously, there exists the odd multipole preference at low multipole ($2\le l\le22$).
Considering the odd multipole preference, we may show the angular correlation is given by:
\begin{eqnarray}
C(\theta)&\approx&\alpha(1-\varepsilon)\,F(\theta)+\alpha(1+\varepsilon)\,G(\theta)\nonumber\\
&&+\sum_{l=23} C_l\,\frac{2l+1}{4\pi}\,W_l\,P_l(\cos\theta),\label{C_FG}
\end{eqnarray}
where 
\begin{eqnarray*}
F(\theta)&=&\sum^{22}_{l}\,\frac{2l+1}{2l(l+1)}\,W_l\,\,P_l(\cos\theta)\:\cos^2\left(\frac{l\pi}{2}\right),\\
G(\theta)&=&\sum^{22}_{l}\,\frac{2l+1}{2l(l+1)}\,W_l\,\,P_l(\cos\theta)\:\sin^2\left(\frac{l\pi}{2}\right),
\end{eqnarray*}
and $\varepsilon$ is a constant related to the parity asymmetry, which is defined to be positive for the odd parity preference and negative for the even parity preference.
Accordingly, $\alpha\varepsilon(-F(\theta)+G(\theta))$ corresponds to the deviation from the standard model, due to the odd multipole preference ($2\le l\le22$).

\begin{figure}[htb!]
\centering\includegraphics[width=0.4\textwidth]{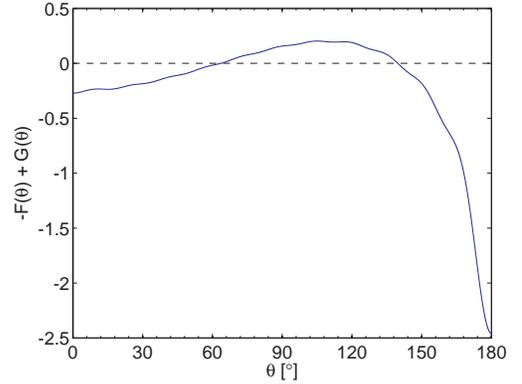}
\caption{the effect of the odd multipole preference on the correlation}
\label{W}
\end{figure}
\begin{figure}[htb!]
\centering\includegraphics[width=0.4\textwidth]{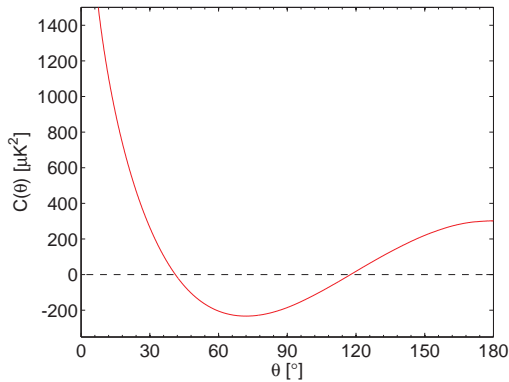}
\caption{the angular correlation without odd-parity preference (i.e. Eq. \ref{cor})}
\label{C_theory}
\end{figure}
In Fig. \ref{W} and \ref{C_theory}, we show $-F(\theta)+G(\theta)$ and the angular correlation of the standard model (i.e. $\varepsilon=0$). 
Let us consider the intervals $60^\circ\le \theta\le 120^\circ$ and $120^\circ\le \theta\le 180^\circ$, which are associated with the statistic $S_{1/2}$.
At the interval $60^\circ\le \theta\le 120^\circ$, the angular correlation has negative values, while the deviation $\alpha\,\varepsilon(-F(\theta)+G(\theta))$ is positive. 
At the interval $120^\circ\le \theta\le 180^\circ$, the angular correlation has positive values, while the deviation $\alpha\,\varepsilon(-F(\theta)+G(\theta))$ is negative. 
Therefore, we find
\begin{eqnarray}
(\left.C(\theta)\right|_{\varepsilon>0})^2<(\left.C(\theta)\right|_{\varepsilon=0})^2\qquad(60^\circ\le \theta\le 180^\circ). \label{C_ne}
\end{eqnarray}
From Eq. \ref{C_ne}, we may see the odd-parity preference (i.e. $\epsilon>0$) leads to the lack of large-angle correlation power.

\begin{figure}[htb!]
\centering\includegraphics[scale=.45]{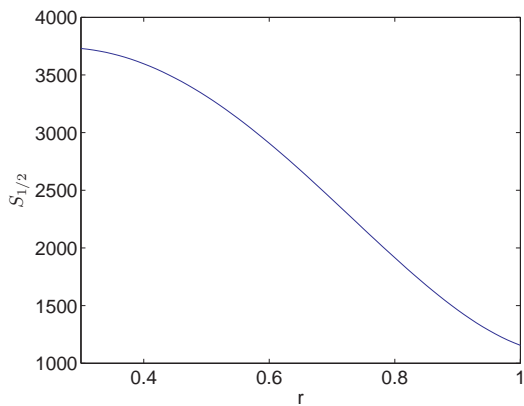}
\caption{$S_{1/2}$ of the WMAP team's Internal Linear Combination (ILC) map, where the octupole components is multiplied by the suppression factor $r$.}
\label{C_l3}
\end{figure}
We like to stress that simple suppression of the power at a single multipole does not necessarily leads to the lack of large-angle correlation.
For instance, suppressing octupole power, which mitigates the odd-parity preference, rather increases the large-angle correlation power.
In Fig. \ref{C_l3}, we show $S_{1/2}$ of the WMAP team's Internal Linear Combination (ILC) map, where we have multiplied the suppression factor $r$ to the quadrupole component of the map. From Fig. \ref{C_l3}, we may see that the value of $S_{1/2}$ rather increases, as the octupole component is suppressed.

\section{Discussion}
\label{discussion}
We have investigated the symmetry and antisymmetry of the CMB anisotropy under the coordinate inversion, which are equivalent to the even and odd parity respectively. As presented in this work, we find there is an anomalous odd-parity preference at low multipole CMB data.
We have investigated non-cosmological origins, and do not find definite association with known systematics.
Among cosmological origins, topological models or primordial power spectrum of feature might provide theoretical explanation, though currently available models do not.
One of a viable phenomenological model requires the real part of the primordial fluctuation be suppressed at low wavenumbers, which leads to violation of translation invariance in primordial Universe on the scales larger than $4\,\mathrm{Gpc}$. 
Additionally, we have compared the phase of even and odd multipole data, and find they show behavior distinct from from each other.

The WMAP power contrast anomaly between even and odd multipoles is explicitly associated with the angular power spectrum data, which are mainly used to fit a cosmological model. Having noted this, we have investigated whether even(odd) low multipole data set is consistent with the WMAP concordance model, and found significant tension. 
We believe these parametric tensions indicate either unaccounted contamination or insufficiency of the assumed parametric model.

Noting the equivalence between the power spectrum and the correlation, we have investigate their association and found that 
the lack of large-angle correlation is phenomenologically identical with the odd-parity preference at low multipoles. 
Additionally, the low quadrupole power may be considered as a part of the odd-parity preference anomaly at low multipoles.

Depending on the type of cosmological origins, distinct anomalies are predicted in polarization data.
Therefore, the upcoming Planck polarization data, which have low noise and large sky coverage, will greatly help us to understand the underlying origin of the anomaly.

\section{Acknowledgments}
We acknowledge the use of the Legacy Archive for Microwave Background Data Analysis (LAMBDA). 
Our data analysis made the use of HEALPix \citep{HEALPix:Primer,HEALPix:framework} and SpICE \citep{spice1,spice2}.  
This work is supported in part by Danmarks Grundforskningsfond, which allowed the establishment of the Danish Discovery Center.

\bibliographystyle{unsrt}
\bibliography{/home/tac/jkim/Documents/bibliography}
\end{document}